\documentclass[acmtog]{acmart}
\acmSubmissionID{312}
\usepackage{enumitem}

\usepackage{booktabs} 
\usepackage{rotating}
\usepackage{wrapfig,graphicx}
\usepackage[caption=false]{subfig}
\usepackage{soul}
\usepackage{svg}
\usepackage{xspace}
\usepackage{mathrsfs}
\usepackage[normalem]{ulem}

\acmPrice{15.00}

\setcopyright{acmlicensed}\acmJournal{TOG}
\acmYear{2023}\acmVolume{40}\acmNumber{6}\acmArticle{1}\acmMonth{12} \acmDOI{10.1145/3478513.3480562}

\setcitestyle{square}

\theoremstyle{definition}

\newif\ifshowauthorcomments

\showauthorcommentsfalse

\newcommand{\geos}{{geometric deformations}}

\newcommand{\Geos}{Geometric deformations}
\newcommand{\topos}{constructive operations}
\newcommand{\Topos}{Constructive operations}

\DeclareMathOperator*{\argmin}{argmin}   

\newcommand{\B}{\mathcal{B}}
\newcommand{\M}{\mathcal{M}}

\newcommand{\stkout}[1]{\ifmmode\text{\sout{\ensuremath{#1}}}\else\sout{#1}\fi}

\newif\ifseechanges

\ifseechanges

\else

\fi

\begin{document}

\title{B-rep Matching for Collaborating Across CAD Systems}

 \author{Benjamin Jones}
 \authornote{Equal contribution}
 \affiliation{%
   \institution{University of Washington}
   \country{USA}}

 \author{James Noeckel}
 \authornotemark[1]
 \affiliation{%
   \institution{University of Washington}
   \country{USA}}

 \author{Milin Kodnongbua}
 \authornotemark[1]
 \affiliation{%
   \institution{University of Washington}
   \country{USA}}

 \author{Ilya Baran}
 \affiliation{%
   \institution{PTC}
   \country{USA}}

\author{Adriana Schulz}
\affiliation{%
  \institution{University of Washington}
  \country{USA}}

\renewcommand{\shortauthors}{Jones et al.}

\begin{abstract}

Large Computer-Aided Design (CAD) projects usually require collaboration across many different CAD systems as well as applications that interoperate with them for manufacturing, visualization, or simulation. A fundamental barrier to such collaborations is the ability to refer to parts of the geometry (such as a specific face) robustly under geometric and/or topological changes to the model. Persistent referencing schemes are a fundamental aspect of most CAD tools, but models that are shared across systems cannot generally make use of these internal referencing mechanisms, creating a challenge for collaboration. In this work, we address this issue by developing a novel learning-based algorithm that can automatically find correspondences between two CAD models using the standard representation used for sharing models across CAD systems: the Boundary-Representation (B-rep). Because our method works directly on B-reps it can be generalized across different CAD applications enabling collaboration.

\end{abstract}



\setcopyright{acmlicensed}
\acmJournal{TOG}
\acmYear{2023} \acmVolume{42} \acmNumber{4} \acmArticle{1} \acmMonth{8} \acmPrice{15.00}\acmDOI{10.1145/3592125}

\begin{teaserfigure}
  \centering
  \includegraphics[width=\textwidth]{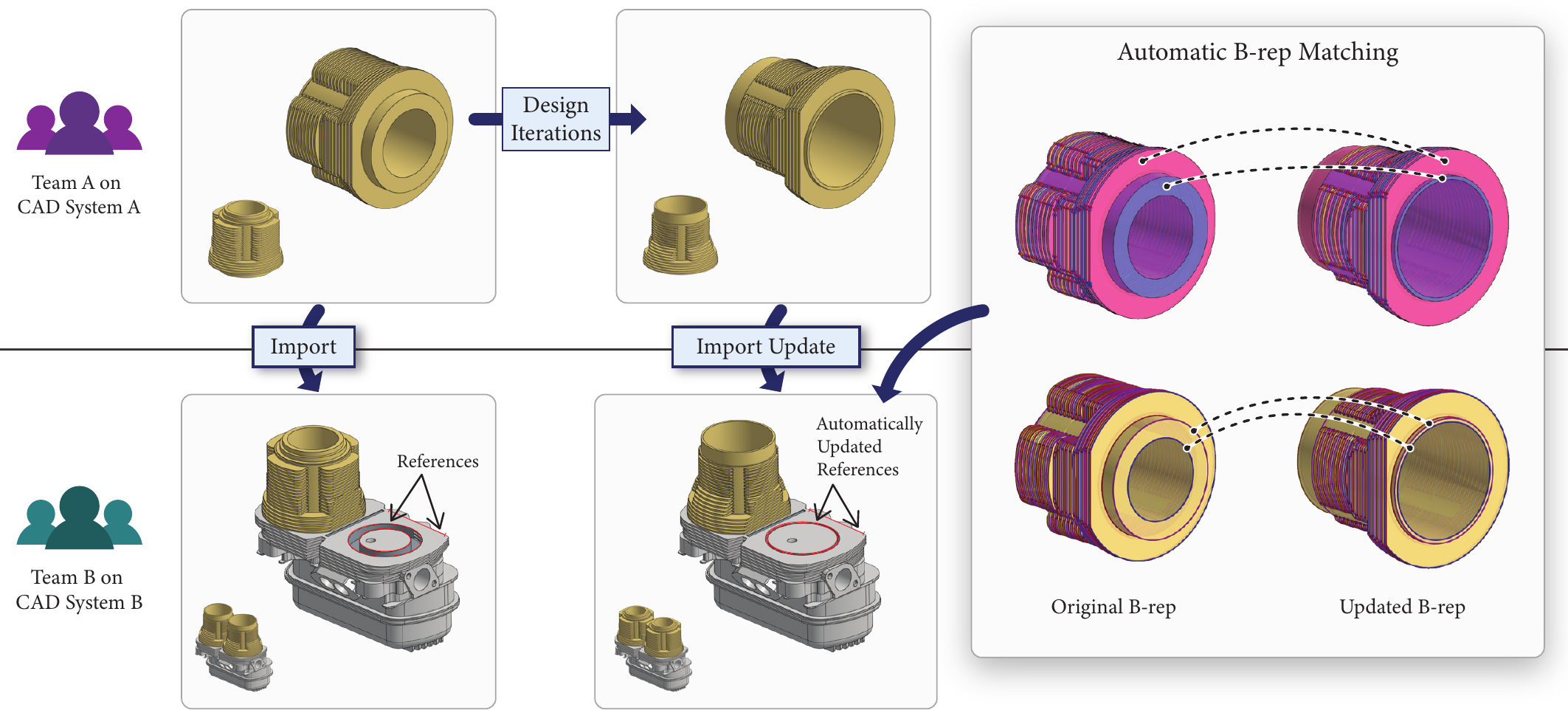}
  \caption{Example of collaborative workflow. Team B is modeling an engine in System B and importing a cylinder that is designed by team A in a different CAD system. The engine's cylinder base geometry is referenced off of the imported cylinder to ensure proper fit. When team A updates the cylinder design, team B imports the new B-rep model and our method enables references to be automatically re-assigned. This enables the CAD model in System B to be automatically updated to match the edits made by user A. In this example, the size and positions of cutouts are updated to align with the new model, and the overall length of the base plate changes, becoming smaller to match the new cylinder dimensions.} 
  \label{fig:teaser}
\end{teaserfigure}

\maketitle
\section{Introduction}
\label{sec:introduction}

Most manufactured objects are designed with commercial Computer-Aided Design (CAD) systems. There is a large number of such systems on the market as well as applications that interoperate with them, including Computer-Aided Manufacturing (CAM) systems, visualization systems, and simulations/analysis systems. Large projects typically require collaboration across several such systems. For example, a team of engineers, B, may design an engine in one CAD system using a cylinder that is designed by team A in another system. When designing the engine, team B would import the cylinder's model into their CAD system and use its features as \emph{references} (see Figure~\ref{fig:teaser}). A fundamental challenge with such collaborations is preserving such \emph{references} after models are updated---e.g, if team B modifies the piston and sends the updated model to team A.

Persistent \emph{referencing} is the ability to programmatically refer to  parts of a CAD model and it is a fundamental construct of modern CAD systems---for example, the chamfer operation references the specific edge where it should be applied. CAD references enable robust design iteration--- for instance, if the CAD program parameters are changed, altering the shape and position of that edge, the CAD system still ``knows'' which edge needs to be chamfered. However, models that are shared across systems cannot generally make use of these referencing schemes, as common CAD exchange formats lack the data structures CAD systems use internally to track entities across edits. This creates a challenge for collaboration, since users would need to manually specify all references every time they re-import a model that has been updated in a different system.

In this work, we address this need by proposing a novel approach to persistent referencing that is agnostic to the CAD system: geometric matching. We observe that what makes CAD referencing schemes system-specific --- inhibiting collaboration --- is that they are unique, proprietary, and rely on the program history: the sequence of CAD operations that construct the shape. The reason why programmatic tracking approaches have been favored over purely geometric models is, of course, the great challenge of robust geometric matching under the wide range of topological variations common in CAD design iterations. Since program-based historical information is available \emph{within} a single CAD system, it can be leveraged to construct heuristic-driven tracking.

Our method takes as input two CAD models: the original $\mathcal{B}_o$ and the updated version $\mathcal{B}_u$. These models are expressed in the standard geometric representation used across all CAD systems: the Boundary-Representation (B-rep). B-rep represents CAD models with infinite resolution, as a topological graph of \emph{entities} (faces, edges, and vertices) each of which has an associated parametric geometry (surfaces, curves, and points, respectively). Our method automatically computes matches between entities of $\mathcal{B}_o$ and $\mathcal{B}_u$. By transferring references across the matched entities, this algorithm allows for seamless collaboration across CAD systems.   
An example use case is illustrated in Figure~\ref{fig:teaser}. In the example, the cylinder is updated and its B-rep is re-imported into the CAD system where the engine is being designed. Our method then automatically finds correspondences between the new geometry and the previous import allowing references to be directly transferred. As a result, the cylinder valve is automatically updated when the CAD program is executed with the updated references, changing overall dimensions and position of cutouts to match the imported geometry.

Our matching algorithm is based on two observations. First, we observe that matches depend not only on the geometric features of B-rep entities but also on the topological similarities---e.g. a model may have many edges that are geometrically similar (straight curves) but they may be easy to distinguish based on the neighboring faces. Second, we observe that CAD model updates tend to affect entities in a non-uniform manner, as some regions of the model may drastically change while others stay intact. As a result, some regions of the model are ``easier'' to match, while others are ``harder''. 

Based on these two observations our key insight is to use an iterative approach that can leverage the ``easier'' matches to find ``harder'' ones from neighborhood information. At each iteration, our algorithm will take a partially matched B-rep pair and suggest the most likely best match to add to the partial matching. Since deterministic algorithms for scoring such matches are challenging to design and lack robustness, we propose to use machine learning for match selection. We bootstrap this algorithm by initializing the partial match with a geometric matching algorithm that searches for entities that are unchanged between the B-reps. This step can be done robustly with a conservative geometric algorithm.

Our proposed framework has three technical contributions. First, we generate a synthetic dataset of B-rep pairs for training and we release this collection for future research. Second, we develop an inference algorithm for scoring partial matches that can be used at any matching stage (number of partial matches). Finally, we develop an end-to-end algorithm for matching B-reps that combines geometric bootstrapping with iterative inference. We evaluate our approach on synthetically generated ground truth, compare it to different baselines, and further evaluate on a smaller expert-generated dataset to validate that our approach is applicable to real CAD workflows.
\section{Background and Related Work}
\label{sec:related work}

We review the state of the art of referencing and B-rep matching in commercial CAD systems, then discuss related research on shape matching and B-rep learning.

\subsection{CAD Referencing}

\begin{figure}[h!]
    \centering
    \includegraphics[width = \linewidth]{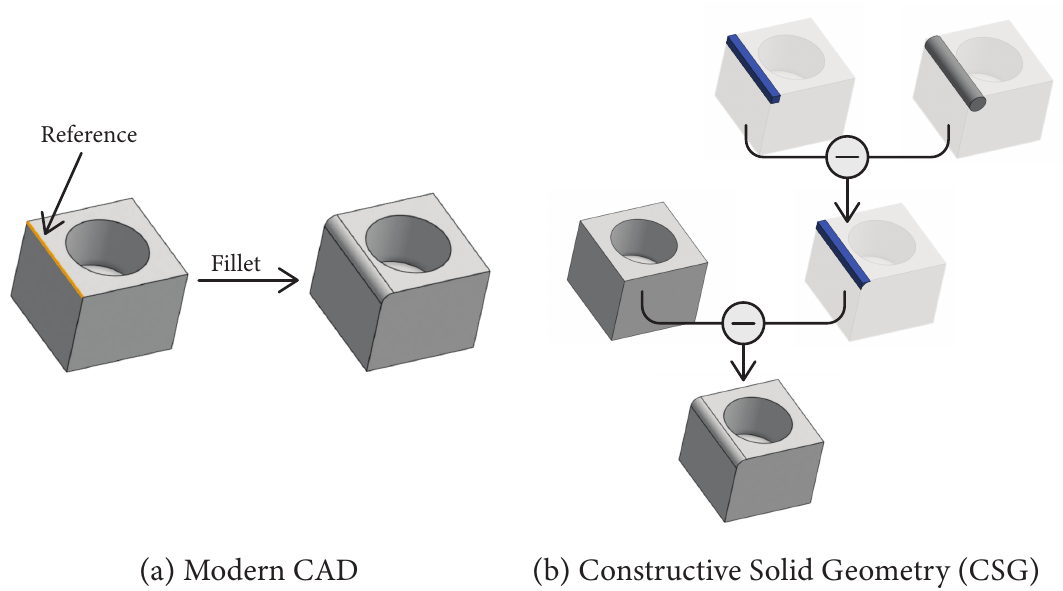}
    \caption{We compare the task of rounding an edge in a modern CAD system (Onshape) and in Constructive Solid Geometry (CSG). While CSG would require a user to manually specify appropriate parameters (translations, scaling) in addition to the sequence of boolean operations, a modern CAD system would allow the user to simply call the \emph{fillet} operation with a \emph{reference} to the edge. This means that the user can change the radius of the fillet by changing only one parameter. Further, and importantly, if the rest of the model is updated (e.g. the cube is resized or the hole is removed), CAD systems can identify the edge and apply the fillet.}
    \label{fig:referencingvcCSG}
\end{figure}

 CAD models are constructed by tens to thousands of CAD operations (called features), most of which use references to intermediate geometry (see Figure~\ref{fig:referencingvcCSG}). Persistent referencing is therefore prevalent in the CAD pipeline and has been the topic of decades of active research both in academia and in industry. Typical approaches fall within two categories. \emph{State-based} referencing schemes allow users to provide custom geometric logic and are common in procedural interfaces (such as CadQuery, Grasshopper, and Houdini). Interactive CAD systems (such as OnShape, SolidWorks, and Fusion360), on the other hand, automatically generate referencing code from user interactions. Such tracking algorithms typically incorporate historical feature information in addition to geometric cues---e.g., a face created by an extrude operation is labeled as the result of that extrude~\cite{farjana2018mechanisms, cheon2012name,bidarra2000semantic, bidarra2005feature}. It is important to note persistent referencing is an inherently ambiguous problem and that typical tracking schemes use heuristics to resolve them, leading to system fragility~\cite{dorribo2016parametric,Thefaile61:online}. That said, by leveraging information from the program history in addition to pure geometric information these algorithms perform significantly better than any heuristic that can be developed without this additional contextual data. The goal of this paper is to bridge the gap between schemes based on programmatic tracking, and geometric matching. Instead of proposing novel heuristics, our approach is to learn from data.

\subsection{B-rep Matching in Commercial CAD Systems}

Since enabling collaboration across CAD systems is a fundamental part of the CAD pipeline, some CAD systems enable some form of matching across systems. A common approach is to use the referencing intelligence of the exporting software system, either via an API connection (which requires both systems installed on the same computer) or by reverse-engineering the proprietary CAD-specific file format. For example, CREO's Unite Technology includes special-purpose methods to enable collaboration with CATIA, Siemens NX, SolidWorks, and Autodesk Inventor.   Though these techniques are proprietary, some are outlined in published patents~\cite{vandenbrande2013methods,spitz2007boundary}. Notably, these approaches rely on specialized data formats and are therefore not general to all CAD systems.

To enable collaboration across any CAD system, some effort has been made to automatically compute correspondences directly on B-reps. Many systems (e.g., Solidworks) can match topological entities that are not modified during the update. A more general method is described in~\cite{vandenbrande2013methods}, but it requires planar faces, cannot handle large modifications of the model, and only produces a rigid motion. Finally,~\citet{kirkwood2018sustained} propose a user-assisted matching tool that uses heuristics to suggest new matches based on user input. To the best of our knowledge, ours is the first fully automatic approach that can automatically match entities without relying on hand-crafted rules and has been shown to perform well through extensive evaluations.

\subsection{Shape Correspondence and Retrieval}
Computing correspondences between two shapes is a well-studied topic in computer graphics. A thorough review of this literature is beyond the scope of this work, so we refer the reader to recent surveys~\cite{sahilliouglu2020recent,deng2022survey}. Typical methods compute a deformation field that aligns a source surface with a target surface. Representations for such fields include sampled points, patches, and implicit or parametric functions. Conversely, the problem we address is to find a set of pairwise matches between discrete sets of B-rep entities (faces, edges, and vertices). It has been shown that methods that simply transfer learned correspondences from more general representations such as point clouds perform significantly worse than methods that learn to match directly on the B-rep entities~\cite{jones:2021:automate} For this reason, our approach works directly with CAD B-reps. 

There is also a body of work on CAD assembly retrieval, which use geometric or topological information to retrieve assemblies, or partial assemblies, similar to a query CAD model~\cite{lupinetti2019content}. A myriad of geometric features (curvature, dihedral angle, and other surface information), along with shape descriptors (shape distribution, 2D projections, angle distribution, spherical harmonics), have been used to find similar CAD models in large databases. In particular, ~\cite{tao2013partial} performs partial retrieval of CAD models represented as B-reps, and identifies useful geometric and topological descriptors that facilitate finding matching regions of CAD models, such as surface parameters and convexity of adjacent faces. However, we are concerned with matches between individual discrete elements of a B-rep, including separate faces, edges, and vertices, rather than whole regions or parts. Moreover, the existing CAD model matching literature is predominantly focused on query-based part retrieval, with a resulting preference for false positives over false negatives. In our setting, the opposite is true; is is more important not to falsely label matches for the purpose of CAD references.

\subsection{Learning from CAD Data}
Algorithms for learning on CAD B-rep representations have seen increasing use, driven in part by the release of large CAD datasets such as the Onshape Public dataset\cite{koch_abc_2019}, the Fusion 360 dataset~\cite{willis_fusion_2020}, and others~\cite{seff2020sketchgraphs}). Another driving factor has been advances in representation learning techniques for CAD-formatted 3D data. Graph neural networks on the topological graphs of the B-rep solids are the prevalent strategy for learning on B-rep data; this is the approach of BRepNet~\cite{lambourne_brepnet_2021}, UV-Net~\cite{jayaraman2021uv}, SB-GCN~\cite{jones:2021:automate}. While the first two approaches learn over a reduced graph of B-rep faces, the latter considers all topological entities (faces, edges, and vertices) and is therefore the architecture we leverage in this work to initially embed our input, with key modifications to account for partially known matches between topological entities.

\paragraph{Data-Driven CAD Applications}
Several methods have been developed to assist users in various aspects of computer-aided design. For example,  AutoMate~\cite{jones:2021:automate} and JoinABLe~\cite{willis2021joinable} learn to assemble CAD models from known parts. \cite{shijie2022material} use graph neural networks to automatically predict correct materials for parts in assemblies, although they do not deal with CAD geometry representations directly, instead using part metadata and rendered images as part-level features.

Another direction that some recent works on CAD learning have taken is to treat CAD B-reps or programs as sequential input, and utilize transformer encoders/decoders to embed or generate entire CAD models. Generative models using transformers have been developed for CAD shapes represented as programs~\cite{wu_deepcad_2021,xu_skexgen_2022} and B-reps~\cite{jayaraman2022solidgen,Guo2022Complexgen}, demonstrating the ability to produce plausible examples of CAD models, albeit with limited scale and complexity.

Other data-driven CAD applications include segmentation~\cite{cao2020graph}, classification~\cite{starly2019fabwave}, and reverse engineering of editable models~\cite{xu2021zonegraphs, rahimi2022reconstructing, uy-point2cyl-cvpr22}. Recent work looks at reducing the size of the dataset necessary to train these models using self-supervision~\cite{jones2022self}. To the best of our knowledge, we are the first to address the issue of finding correspondences between an original and an updated version of a B-rep. 

\begin{figure*}[h!]
    \centering
    \includegraphics[width=\linewidth]{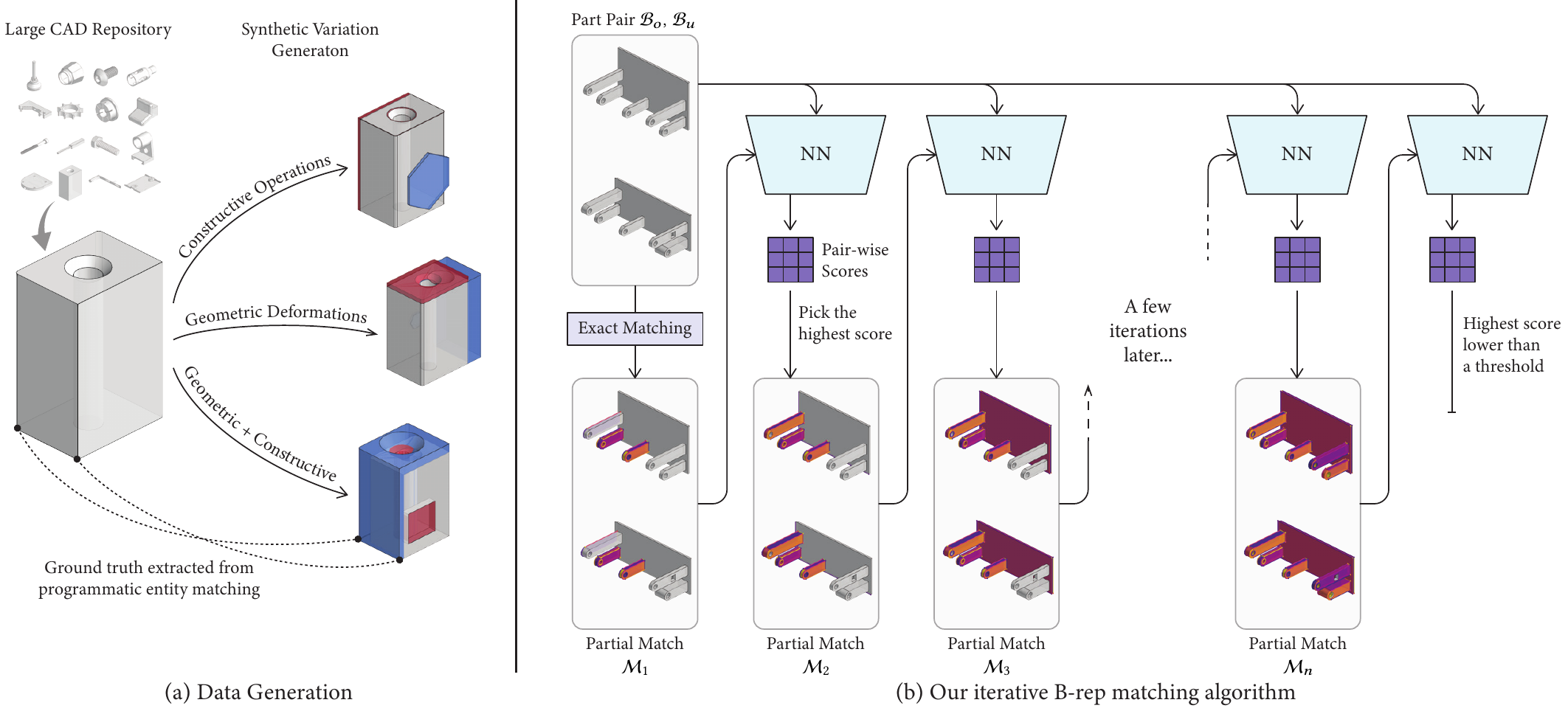}
    \caption{(a) Overview of our methods for data generation of B-rep matching. We draw human-created models from a large public CAD repository, and programmatically apply constructive operations and geometric deformations to create part variations. Because these edits are applied programmatically within the CAD system, we can use its programmatic entity tracking to extract ground-truth matches. (b) We use this data to train a neural network that scores potential entity matches given the geometry and topology of two B-reps and a partial matching. Our matching algorithm initializes the partial matching with exact geometric matches, then iteratively selects the next most likely match then re-evaluates matching likelihoods up to a user specified threshold. }
    \label{fig:overview}
\end{figure*}

\section{Overview}

The goal of our work is to allow references to be propagated from different versions of CAD models once they are exported into the common sharing formats used when collaborating across CAD systems. This format, the B-rep, represents the geometry as a graph of entities (including faces, edges, and vertices). Graph nodes are associated with parametric equations that define their associated geometry (surfaces, curves, and points, respectively). Graph edges denote boundary relationships; vertices bound edges, and closed \emph{loops} of edges bound faces. Since loops aggregate other geometry, they do not have an associated parametric definition, but instead a label denoting if they are an inner or outer boundary of the face they bound. Our task is to search for pairs of matched primary entities --- faces, edges, and vertices --- across the two B-reps. An overview of our system is depicted in Figure~\ref{fig:overview}.

Maintaining references across multiple CAD environments is challenging since imported models contain only geometric information, lacking the CAD program history normally used to track entities that change due to edits. One of our key observations is that while history-based tracking cannot be replicated in our context, it can instead be \emph{learned}. 
We propose to use this logic to generate a large dataset of matches with ground truth labels. By analyzing a smaller set of expert-designed pairs of models and their variations, we identify the key variations that happen across versions: \geos{} and \topos{}. We develop algorithms for automatically generating these kinds of variations within a CAD system and use its tracking mechanism to generate matching labels.

Our method for learning to match over this dataset is designed around the insight that B-rep entity correspondence is informed not only by the geometry of an entity but also by its neighborhood information. This indicates that when some matches are known, they can be leveraged to find novel matches. Our method, therefore, starts by first taking advantage of geometric information to find and match geometrically equivalent B-rep entities---i.e., entities that were not altered between versions. We then use this matching to bootstrap an iterative process that takes advantage of known matches to consecutively tackle increasingly difficult matches. We use a graph neural network to embed the geometric and topological information of entities in both B-reps, and combine this with a partial match to score all potential next matches, from which we greedily choose the next most likely match. We then iteratively re-score the remaining matches using the updated match prior, and continue until a likelihood cutoff is reached.
\section{Data Generation}
\label{sec:datacollection}


\begin{figure*}[t!]
    \centering
    \includegraphics[width=\linewidth]{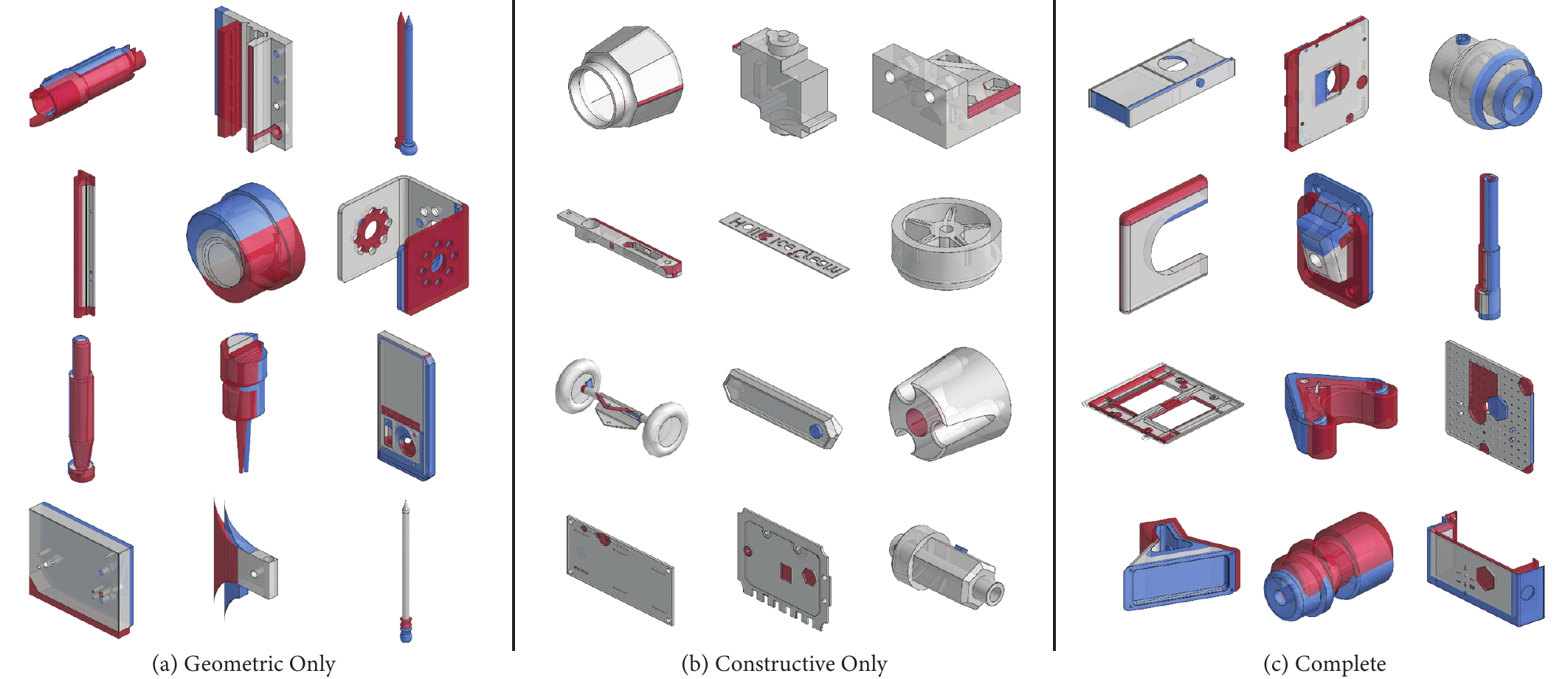}
    \caption{(a-c) Examples of \geos{}, \topos{}, and our complete synthetic dataset, respectively. Red and blue depicts the exclusive regions of original and updated B-reps, respectively, and gray depicts shared regions between the two.}
    \label{fig:datasetOverview}
\end{figure*}

To train and validate our data-driven matching algorithm, we collect models using Onshape's public data repository and scripting capabilities, and either generate or find variations representative of typical CAD revision updates.

\paragraph{Expert Collection}
To validate our method on real-world use cases, we collected examples of updated imports---instances where a B-rep was imported first into Onshape from a different source and the content was then updated to a new version---from Onshape's public repository. 
While this scenario matches directly to our target application, such examples are challenging to find because doing so requires a partially manual search. This workflow is also more common in Onshape enterprise users (companies that work on large collaborative projects) and those models are not made public. 

However, since Onshape has a built-in version control system, examples of CAD revisions are easily accessible for every model in their collection by looking backward in the history of a model. Because only certain points in the revision history correspond to versions that are ready to be exported (for example, finished versions versus edits-in-progress), we manually searched the collection with the help of a CAD expert to identify version pairs representative of typical CAD workflows.

Using these two methods, we compiled a collection of 25 models: 5 from re-imports and 20 from version control revisions. Although this dataset lacks ground truth labels, and is not large enough to train over, it is a representative sample of our target application that can be used to validate our algorithm. We further take inspiration from the variations observed in these examples in constructing our synthetic training dataset.

\paragraph{Synthetic Variation Collection}
We built a collection of synthetic variations of human-designed CAD models by scraping 2,400 public models of varying complexity from Onshape's public repository, then automatically applying variations similar to those we observed in the expert validation set. We develop our variation algorithm within the scripting environment of a CAD system. This allows us to make use of internal mechanisms for entity tracking based on program history to create ground truth matches.

By analyzing our expert collection we observe that variations fall within two categories which we term \topos{} and \geos{}. We developed custom operations that automatically generate variations of each type.  

\Topos{} add or remove material from the model---for example, making a hole for a screw, filleting (rounding) an edge, or adding more detail by sketching and extruding some new geometry. While these operations can modify the existing geometry (e.g., a face becomes shorter as its adjacent edge is filleted), these modifications tend to be minor. There is a much greater impact on the model topology: new entities are created, some are removed, and neighborhood information is fundamentally changed. Our custom \topos{} select faces and extrude polygonal sketches or circles to add or remove material. They can also fillet or chamfer edges. To achieve diverse variations that resemble the operations in the expert collection, we use a biased randomization to select the entities on which to apply these operations as well as the ranges for parameters of the operations that depend on model dimensions. 

\Geos, on the other hand, directly alter the geometric properties of existing elements. These typically correspond to parametric variations such as changing the length of an extrusion or modifying a sketch. While the topology can locally change under these variations, it can also be preserved, while the shape and position of entities can vary greatly.  Since finding the right CAD parameters to tune without breaking the models is often challenging if they are not exposed by the designer, we instead resort to direct editing functionality that consists of moving around groups of entities. We use several heuristics to group parts of the model in direct editing move operations, then run validations to discard changes that result in undesirable changes (e.g., models being divided into two parts). We again use biased randomization when selecting which faces to move and how to move them.

Our complete data set is generated by first introducing \geos{} and then  \topos{} over the resulting variations. We generate 4-18 variations selected from a random distribution biased by the size of the model. We also run ablations of our method over a data set of exclusively \geos{} and one with exclusively \topos{}. Figure~\ref{fig:datasetOverview} shows examples of all three sets.  We have released our code on Onshape's public library for reproducibility. 
\section{B-rep Matching Algorithm}

At its core, our method is an iterative greedy matching of two B-rep graphs, $\mathcal{B}_o$ and $\mathcal{B}_u$, to obtain an element-wise matching $\mathcal{M}$ between their topological entities. The basic algorithm is defined by Equation~\ref{eq:iteration}: at each iteration we add the highest probability matching pair from unmatched candidate pairs $C$ of similar topological type (faces with faces, edges with edges, etc.)

\begin{equation}
    \M_{n+1} = \M_n \cup \max_{(i,j) \in C} p_\theta(i,j \;|\; \mathcal{B}_o, \mathcal{B}_u, \mathcal{M}_n)
    \label{eq:iteration}
\end{equation}

Crucially, this probability $p$ is conditioned not only on the topology and geometry and the B-reps, but also on the previous partial matching $\M_n$. Because typical CAD edits leave some portion of the model unchanged, we are able to use exact geometric matching to effectively bootstrap this method. For the other entities, we propose to learn $p_\theta$ from data, the synthesis of which is described in Section~\ref{sec:datacollection}. We continue this iteration until $\max_{(i,j)} p$ is below a user-specified threshold; this allows us to control the precision-recall trade-off at run-time.

\subsection{Learning Matchings from Data}

The scoring function $p$ at the core of our iterative method needs to be able to pick the next most probable match given any particular partial matching $\mathcal{M}_n$. Given a dataset $D = \{(\B_o^0, \B_u^0, \M^0), \ldots \}$ of variation pairs and ground truth matchings, we want to find $p_\theta$ that minimizes the error over all partial matchings:
\begin{equation}
    \min_\theta \sum_{(\B_o, \B_u, \M) \in D} \sum_{\overline{\M} \subset \M} \frac{1}{|\overline{C}|} \sum_{(i,j) \in \overline{C}} \mathcal{L}\left(p_\theta(i,j \;|\; \B_o, \B_u, \overline{\M}), \mathcal{M}(i, j)\right)
\label{eq:minimization}
\end{equation}
where $\overline{\mathcal{M}}$ is one potential partial matching, $\overline{C} = \B_o \otimes \B_u \setminus \overline{\mathcal{M}}$ is the set of candidate unmatched pairs relative to $\overline{\mathcal{M}}$, and $\mathcal{M}(i,j)$ denotes whether $(i,j) \in \mathcal{M}$.\footnote{$\otimes$ denotes a Cartesian product of graph nodes limited to pairs of like topology (faces $\times$ faces, etc.).} 

For our application, match precision is more important than recall. When a match used as a reference to a CAD operation is missed, the CAD program will throw an error on that operation and label the error as a missed reference. If an entity is incorrectly matched, the operation may go through, causing downstream errors that are more difficult to debug. Therefore, we employ a weighted binary cross-entropy loss that gives more weight to negative match examples:
\begin{equation}
\mathcal{L}(\hat{p}, p) = - \left[ p \cdot \log (\hat{p}) + w \cdot (1 - p) \cdot \log (1 - \hat{p}) \right],
\label{eq:lossfunction}
\end{equation}
where $w$ is the weighting constant. Higher $w$ means fewer errors but with the cost of fewer predicted matches. We choose $w = 2$ in our experiments.

In practice, training over every possible partial matching is infeasible, so we estimate $\theta$ as

\begin{equation}
     \argmin_\theta \sum_{k=0}^N \sum_{(\B_o,\B_u,\M)\in D} \frac{1}{|\widetilde{C}|} \sum_{(i,j) \in \widetilde{C}_k} \mathcal{L}\left( p_\theta ( i,j \;|\; \B_o, \B_u, \widetilde{\M}_k), \M(i,j) \right)
\label{eq:trainingapprox}
\end{equation}

Here $\widetilde{\M}_k$ is a partial match consisting of 50\% of the ground truth match $\M$ chosen randomly in the $k$th training epoch.
We train for $N = 1000$ epochs and select $\theta$ by minimum loss on a held out validation set with random but fixed partial matches.

\subsection{Network Architecture}

Our match scoring function $p_\theta(i,j \;|\; \mathcal{B}_o, \mathcal{B}_u, \mathcal{M})$ is conditioned on the geometric and topological structure of the two B-reps, $\mathcal{B}_o$ and $\mathcal{B}_u$, as well as any prior information we have about correspondences between them $\mathcal{M}$. We encode the geometric and topological information of B-rep entities using a hierarchical graph convolutional network, Structured B-rep GCN (SB-GCN)~\cite{jones:2021:automate}. This takes as input both the parametric definitions of geometry, numerically computed statistics about each entity (such as bounding box, surface area, and center of mass), and the topological B-rep graph, and outputs an embedding vector for each B-rep entity. In our experiments, we use a 6-layer SB-GCN with a 64-dimensional embedding space. 

\begin{figure}
    \centering
    \includegraphics[width=\linewidth]{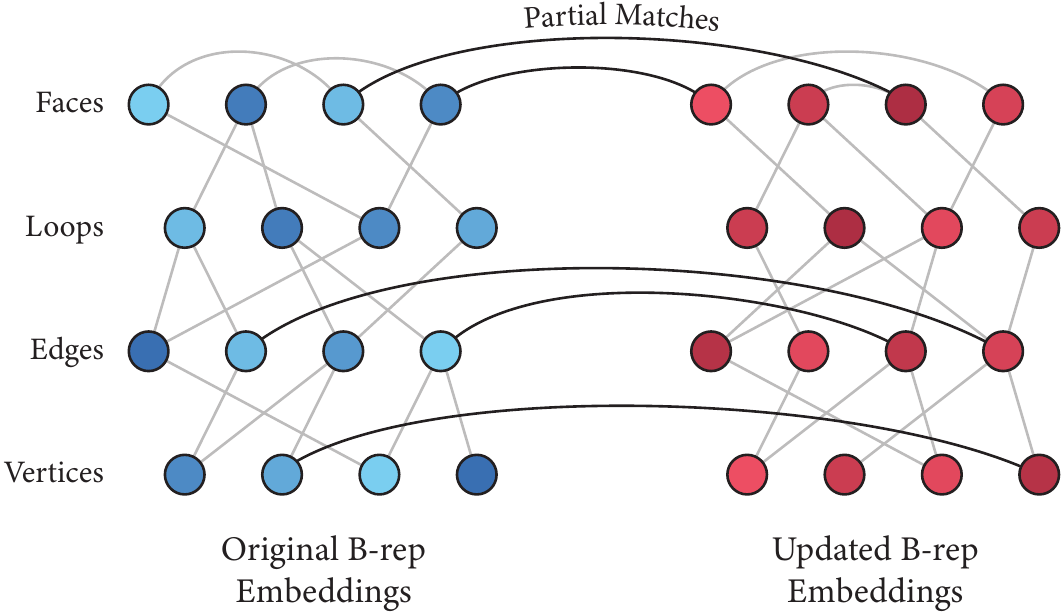}
    \caption{Partial match conditioning for the GAT. Partial matches are represented as cross-graph edges in unification of the B-rep graphs $\B_o$ and $\B_u$.}
    \label{fig:matchconditioning}
\end{figure}

\begin{figure}
    \centering
    \includegraphics[width=\linewidth]{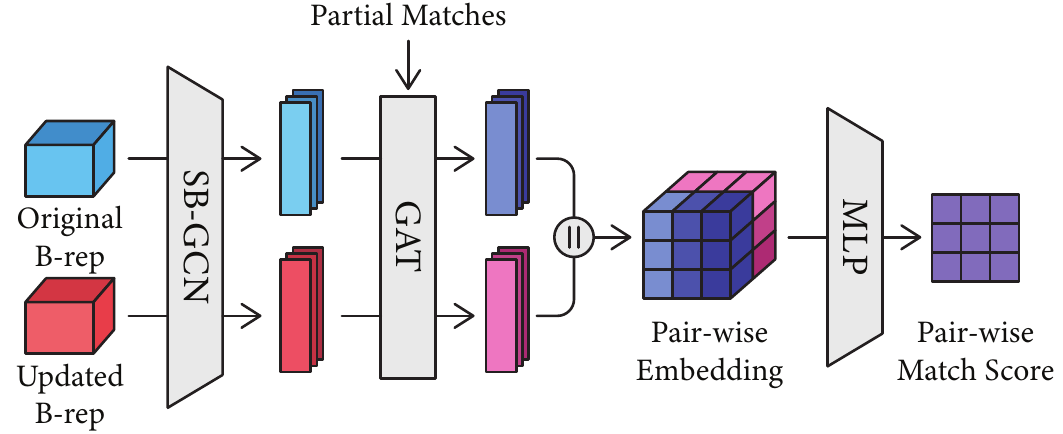}
    \caption{Scoring network overview. Given the original and updated B-reps, the SB-GCN computes the embeddings for each topological entity. The GAT takes partial matches and do message passing on the embeddings across two B-reps. The two sets of embeddings are pair-wise concatenated and passed to an MLP to get pair-wise match scores.}
    \label{fig:scoringnetwork}
\end{figure}

To further condition on prior matching information, we use these embeddings as node features in a second graph convolutional network that performs message passing both within and between the parts. To achieve this, we construct a joint-part graph, encoding both the internal topology of both B-reps, as well as matches between entities of the same type as cross-part edges (see Figure~\ref{fig:matchconditioning}). Node embeddings are initialized as SB-GCN embedding vectors. Edges are undirected, and carry a one-hot encoded type, specifying them as vertex-edge, edge-loop, loop-face, face-face, or prior-match. We compute our final node embeddings using a Graph Attention Network (GAT) with additive edge messages~\cite{velickovic2018graph}. In our experiments we use a 4-layer GAT with 8 attention heads per layer and a 64-dimensional embedding space. Embeddings of candidate pairs are embedded as input to a 2-layer MLP to produce matching score logits. The full scoring architecture is illustrated in Figure~\ref{fig:scoringnetwork}.

\subsection{Bootstrapping Matches}
\label{sec:exact-matching}
Our scoring model is conditional on matching priors. These have two sources, exact geometric matches, and predictions from previous iterations of our model.

\begin{figure} [h!]
    \centering
    \includegraphics[width=.8\linewidth]{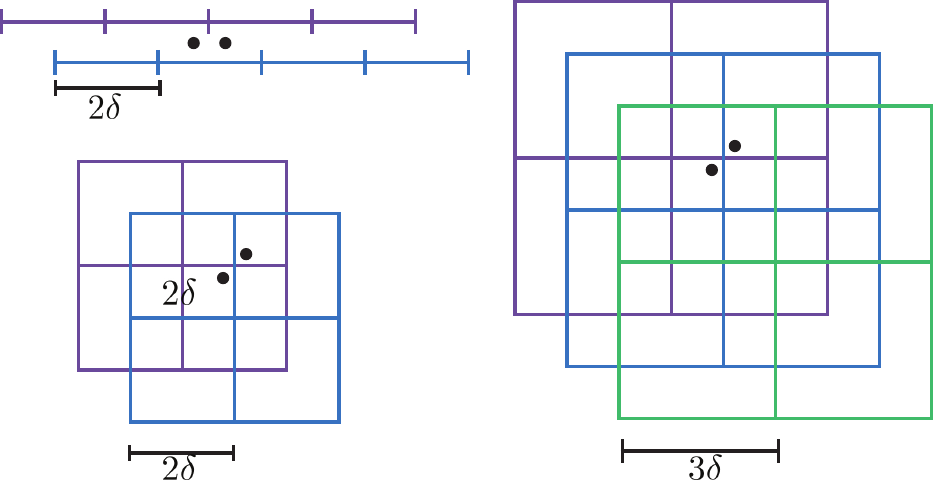}
    \caption{We illustrates our hashing approach using shifted grids. The top left corner illustrates the grids in one dimension for finding solutions that are close up to tolerance $\delta$. In this case we have two grids where each cell has length $2\delta$ and the cells are shifted by $\delta$. Notice that if two points are close within tolerance $\delta$, there will \emph{always} exist one cell either on the top or the bottom grid that contains both points.  To expand this to higher dimensions vectors, we must create $D+1$ grids, of edge length $(D+1)\delta$, where $D$ is the number of dimensions. We illustrate an example of shifted grids for two dimensions (right) and illustrates why simply using two shifted grids does not generalize for higher dimensions (bottom left). While this method deterministically ensures that pairs of points within tolerance are found, it may further return points that are not within tolerance. Specifically, it may return points that are at most $(D+1)\sqrt{D}\delta$ apart (the dimensions of the diagonal of the grid cell.) In our solution we further check those points and filter them out.}
    \label{fig:lsh}
\end{figure}

As previously discussed, unchanged  entities are common in CAD design updates since iterations tend to affect only sections of the model. We can use a B-rep kernel (e.g. Parasolid) to check if two entities are coincident. To speed up processing, rather than checking every pair, we only check the models whose centroids match, which can be checked for efficiently by hashing the 3D vector representing the centroid of each topological entity in the old body. To account for tolerances, we propose a 3D nearest neighbor structure using shifted grids, which is efficient in low dimensions and deterministic (see Figure~\ref{fig:lsh}).

\section{Results}
\label{sec:results}

Our method was trained using our synthetic data.
The training took 12 hours on average using a single NVIDIA RTX 2080 Ti. The trained network was evaluated against the generated ground truth and we further evaluate our approach over the expert data set to validate that our method, trained on synthetic data, generalizes well to real-world data.

\subsection{Synthetic Data Evaluation}

We constructed a synthetic variation dataset using the \geos{} and \topos{} described in Section~\ref{sec:datacollection}. We collected 2,400 CAD models and generated 3 variations of each. After filtering parts that caused import errors and scale outliers over 100 times the size of the median part, our dataset contains 2,266 original models and 6,257 variations. We split this dataset by original model 80-10-10\% for training, validation, to ensure both models remain unseen between training and testing. We report the results of our approach and compare them to deterministic baselines and ablations. We refer readers to the table in the supplemental for numerical comparisons between methods.

\paragraph{Results Across Different Entity Types}

Our method correctly matches $91.5\%$, $92.1\%$, and $94.9\%$ of faces, edges, and vertices, respectively. We note that not all entities have ground truth matches because entities can be added or deleted over design iterations. Our goal is to match as many entities as possible while avoiding errors. We therefore evaluate our results by categorizing the entities in $\mathcal{B}_u$ into five groups:
\begin{itemize}[leftmargin=*]
    \item \emph{True Positive.} Entities from $\mathcal{B}_u$ that had a ground truth match in $\mathcal{B}_o$ and were matched correctly by our algorithm.
    \item \emph{True Negative.} Entities from $\mathcal{B}_u$ that did not have a ground truth match in $\mathcal{B}_o$ and were labeled as unmatched by our algorithm.
    \item \emph{Missed.} Entities from $\mathcal{B}_u$ that had a ground truth match in $\mathcal{B}_o$ but were labeled as unmatched by our algorithm
    \item \emph{Incorrect.} Entities from $\mathcal{B}_u$ that had a ground truth match in $\mathcal{B}_u$ but were matched by our algorithm to a different entity in $\mathcal{B}_o$.
    \item \emph{False Positive.} Entities from $\mathcal{B}_u$ that had no ground truth match but were matched by our algorithm.
\end{itemize}

Table \ref{tab:resultSynthetic} reports the percentage of total entities in $\mathcal{B}_u$ that fall within each category. On average, the exact matching finds 15.7\%, 23.8\%, and 35.4\% of the total matches for faces, edges, and vertices. Our neural network finds an additional 72.6\%, 63.9\%, and 55.2\% of the total matches, while maintaining a 2.8\%, 1.9\%, and 1.1\% error rate.

\begin{figure}[h!]
    \centering
    \includegraphics[width =\linewidth]{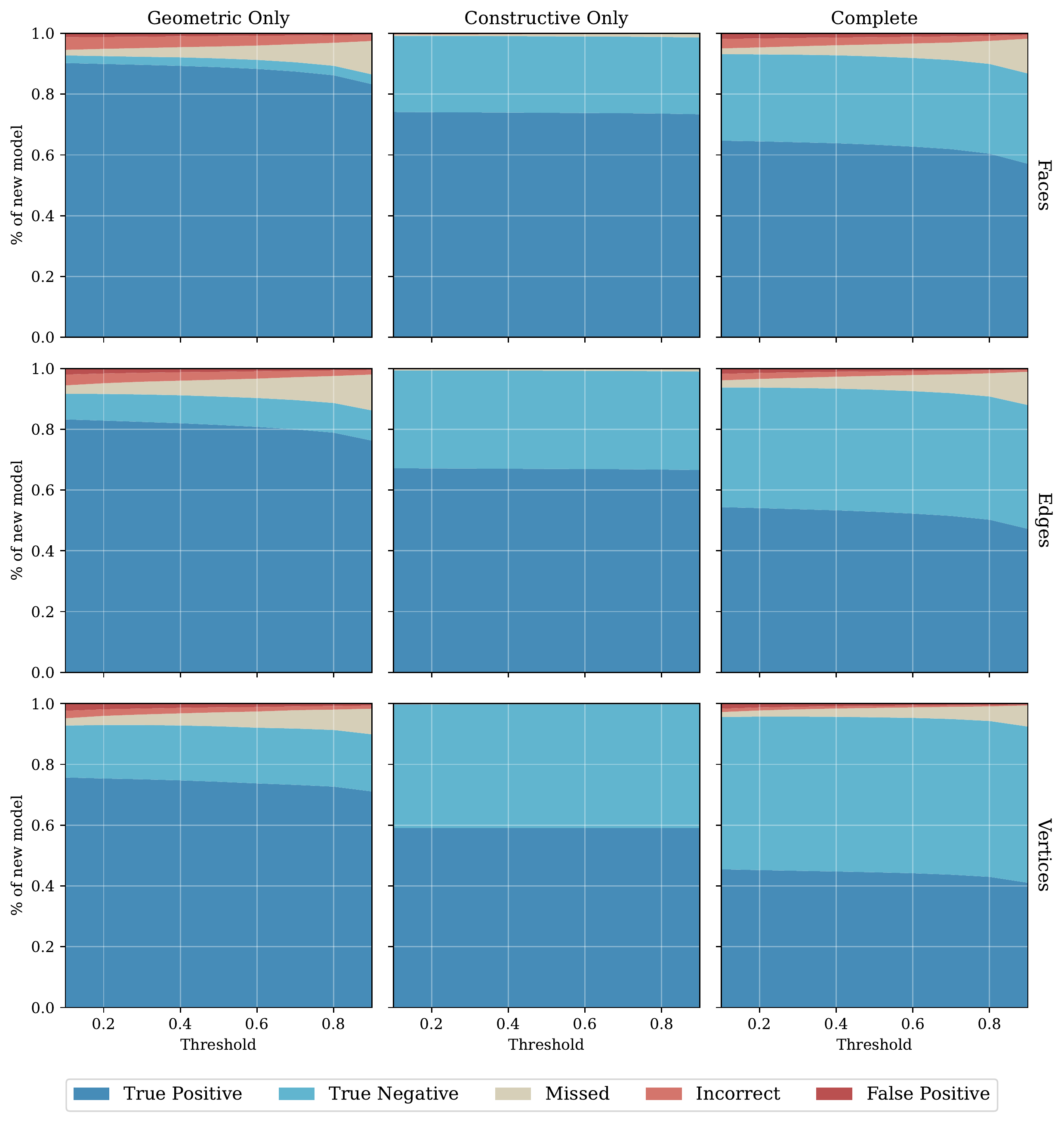}
    \caption{Metrics breakdown at different thresholds on synthetic data}
    \label{fig:ourresult}
\end{figure}

In Figure~\ref{fig:ourresult}, we report the breakdown over different thresholds used as the stopping criteria for the iterative matching algorithm. We stress that for our application, entities that are matched but should not have been matched (both incorrect or false positives, in shades of red) create greater issues for users than missed matches (in beige) as they are more difficult to debug. 

As seen in the third column, which is the result using the complete synthetic data, our method is effective at trading off errors and the number of correct matches as this threshold varies and we have manually selected a threshold of 0.7 based on this result (reported in Table~\ref{tab:resultSynthetic}).  We note that while other thresholds would result in even fewer errors, the 1-3\% error rate already makes these occurrences very infrequent. Further, not all entities are used for references, so when composed with the likelihood of it being used, the chance of these errors causing issues for the user will be even lower. Finally, we note that zero error is not a realistic goal because of potential ambiguities in the ground truth data and potential one-to-many matches we have not accounted for.

The first two columns of Figure~\ref{fig:ourresult} show our model when trained and evaluated on synthetic data generated by two simplifications of our variation algorithm: one using exclusively \geos{} and the other using exclusively \topos{}. We compare these results with the complete synthetic data. We observe that constructive operations lead to a much larger number of true negative, which is to be expected as those operations create new entities that do not have a previous match. They are also significantly easier to match since geometric changes are small. Geometric deformations, on the other hand, create more challenges for matching. Results over the complete dataset have a similar number of true negatives as constructive operations because it incorporates those changes, and also similar to geometric deformations, it has to trade-off missing to incorrect labels because of the challenges created by geometric changes.

\begin{table}[h!]
\small
\caption{Our results on complete synthetic data reported at threshold 0.7.}
\begin{tabular}{lrrrr}
\toprule
& & \multicolumn{1}{c}{Faces} & \multicolumn{1}{c}{Edges} & \multicolumn{1}{c}{Vertices} \\
\midrule
Exact & True Positive & 11.0\% & 14.0\% & 17.1\% \\
\cmidrule(lr){1-5}
\textbf{Ours} & True Positive & 62.0\% & 51.6\% & 43.8\% \\
& True Negative & 29.5\% & 40.5\% & 51.1\% \\
& Missed & 5.6\% & 6.0\% & 4.0\% \\
& Incorrect & 2.0\% & 1.2\% & 0.5\% \\
& False Pos & 0.8\% & 0.7\% & 0.6\% \\
\cmidrule(r){2-5}
& Correct Label & 91.5\% & 92.1\% & 94.9\% \\
& Incorrect Label & 2.8\% & 1.9\% & 1.1\% \\
\bottomrule
\end{tabular}
\label{tab:resultSynthetic}
\end{table}

\paragraph{Deterministic Baselines}
We consider three deterministic baselines which incrementally increase the number of potential matches by trading-off potential errors: coincidence matching, overlap matching, and adjacency propagation. The coincidence matching baseline enables us to evaluate the lift that our network has over exclusively applying the coincidence matching algorithm described in Section~\ref{sec:exact-matching}. 

\begin{figure}[h!]
    \centering
    \includegraphics[width =\linewidth]{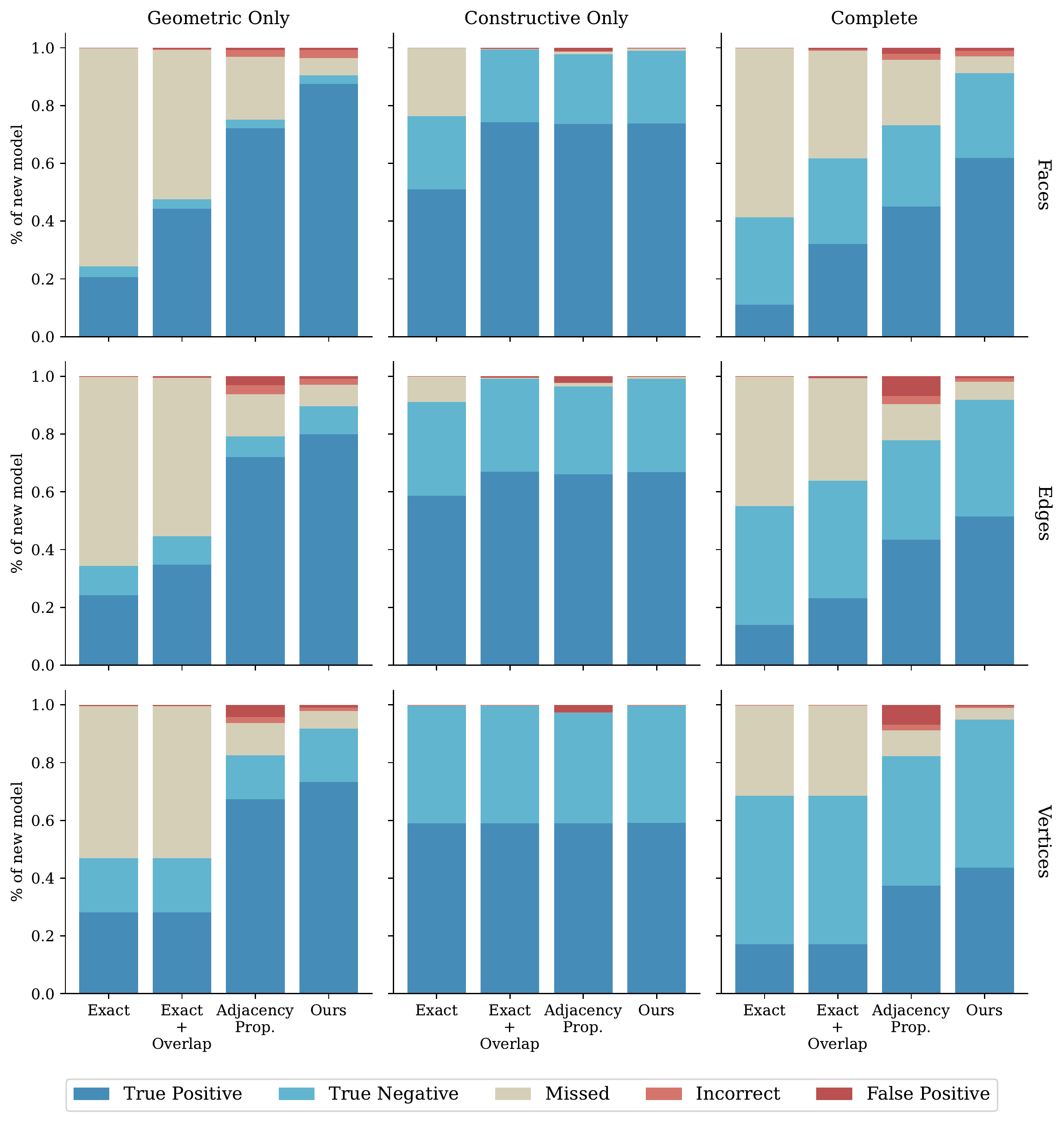}
    \caption{Comparisons with deterministic baselines on synthetic data}
    \label{fig:deterministicComparisons}
\end{figure}

\begin{figure}[h!]
    \centering
    \includegraphics[width=\linewidth]{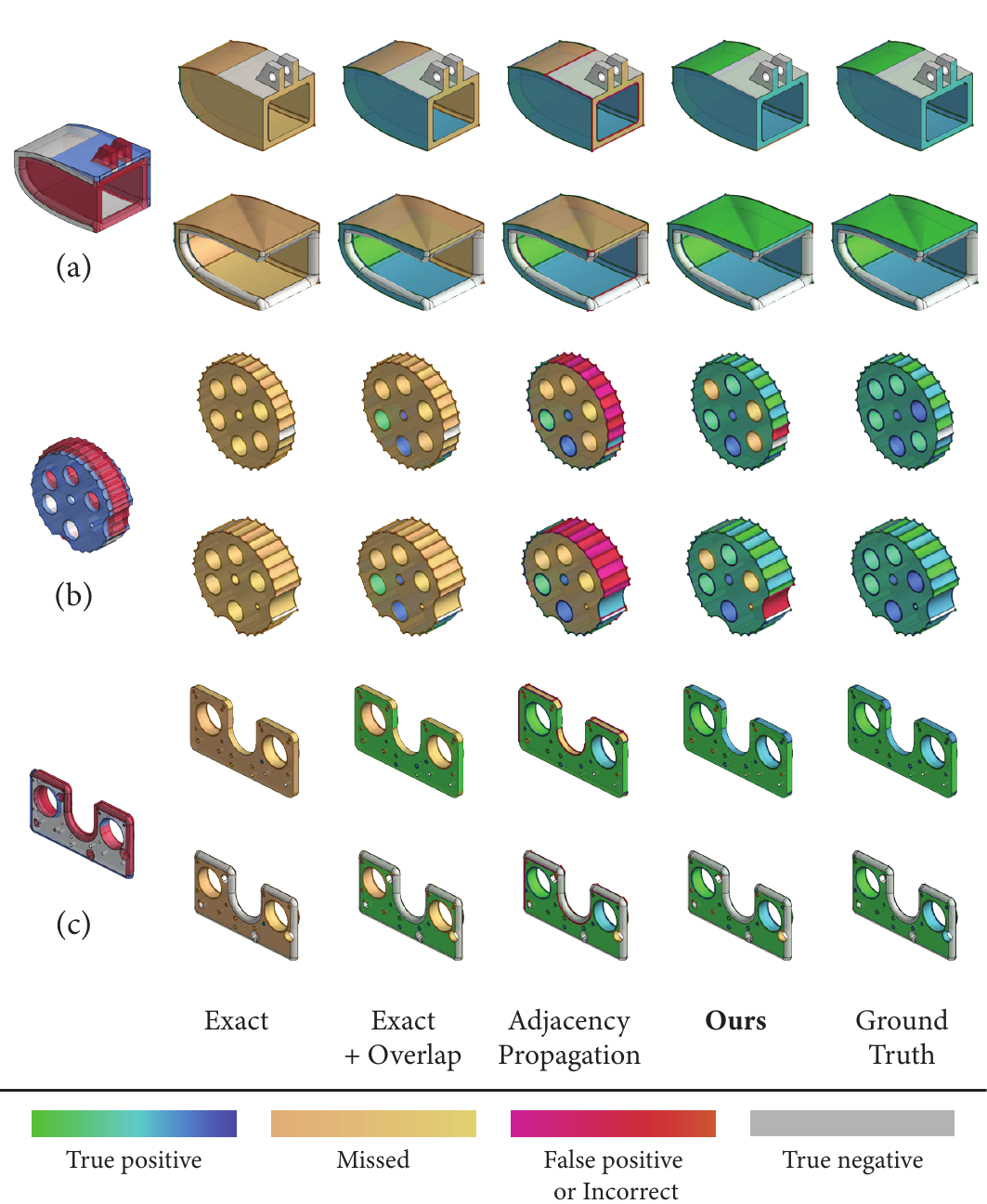}
    \caption{Gallery of B-reps pairs matched by different deterministic baselines, as compared with our method. The first column shows the difference between original and updated B-reps. The remaining columns shows matching performance by different algorithms. For each pair, the top is the original and the bottom row is the updated B-rep. Entities (top and bottom) that were matched are rendered with the same color. True positives are in shades of blue-green; incorrect labels (incorrect and true negatives) in shades of pink-red; missed matches in shades of yellow; and true negatives in gray.}
    \label{fig:syntheticViz}
\end{figure}

There are many cases where the coincidence matching algorithm will fail to match entities with minimal edits to the geometry. These are very common under \topos{}: for example, a face whose edge has been filleted or where a hole has been applied will only reduce in size by a small amount but the underlying parametric geometry will remain unchanged. This observation inspires our second baseline, which matches both entities that are coincident, and ones that have a significant overlap. Our overlap matching function looks for pairs of edges and faces that were not matched by coincidence matching and whose associated geometry (curves and surfaces) are identical. For each pair, we then compute their overlap, and if the overlap is a large enough fraction (80\% in our implementation) of the smaller of the two entities, the pair is considered matched. This computation is accelerated by hashing common geometric types and using the shifted grids algorithm described in Figure~\ref{fig:lsh}. 

Finally, we consider an iterative solution that matches over deterministic signatures of adjacency. This method works by first matching by coincidence and overlap, then iteratively matching entities that have a matching adjacency signature. To create an adjacency signature for a given entity, we traverse its neighborhood graph in counterclockwise order and store the match index of each entity in the neighborhood (using a special value of -1 if that entity is unmatched). We use lexicographical sorting to avoid the ambiguity of which entity to start with, treating loops individually and aggregating them in lexicographic order. Adjacency signatures are computed for entities in $\mathcal{B}_u$ and put in a priority queue based on the ratio of matched to unmatched entities (so that the function matches based on more matching adjacency first). We go over this priority queue creating matches if (a) there is a unique entity if $\mathcal{B}_o$ with the same adjacency signature, or if these can be disambiguated by the geometric heuristics (like preferring to maintain curve/surface types or matching coaxial cylinders). The adjacency signatures are updated every time a match is found and the algorithm terminates when no more matches can be found. This final baseline describes our best attempt at addressing this problem in a deterministic manner. This algorithm has been integrated into the Onshape software and is currently executed every time an Onshape user re-imports a model. See more details in the published patent~\cite{baran:2020:brepmatching}.

We compare our solution with these three deterministic methods in Figure~\ref{fig:deterministicComparisons}, showing results over different entities and testing sets, and in Figure~\ref{fig:syntheticViz} for visualizations. Our method significantly outperforms all baselines. In the complete synthetic data set, when comparing our method with exact matching, we see a lift of 50.2, 37.1, and 26.3 percentage points of correctly labeled faces, edges, and vertices (both shares of blue).

We also observe a large lift over overlap matches: 29.8 and 28.3 percentage points of correctly labeled faces and edges, respectively. We note that such lift is significantly smaller if we do not consider \geos{}, as overlap matches will handle the majority of \topos{} (See the second column of Fig. \ref{fig:deterministicComparisons}). 

Finally, when compared to adjacency propagation, our method not only improves the number of correct labels but is also significantly more robust. For example, in Fig.~\ref{fig:syntheticViz}b, our method was able to correctly match most of the gear teeth while the adjacency propagation method failed to distinguish between them because they are topologically the same. Another common pattern is that the adjacency propagation method will match one the filleted edges (See Fig.~\ref{fig:syntheticViz}a and c). Our method learns not to match these one-to-many cases.

\paragraph{Contrastive Learning Baseline} In addition to these deterministic baselines, we compare against a contrastive learning representation (CLR)-based method that computes matches by greedy matching of per-element embeddings produced by SBGCN, up to a minimum cosine similarity threshold. We use an N-tuplet contrastive loss to encourage matching elements to have similar embeddings. With this method, we find that there exists no similarity threshold that provides a good tradeoff between the number of missed matches and the amount of error. Compared to this baseline, our method has a lift of 48.5, 35.4, and 25.8 percentage points of correctly labeled faces, edges, and vertices. 
Despite attempting many variations of this approach with different loss functions and geometric features presented to the encoder, such as UV-Net embeddings of faces and edges, we were ultimately unable to improve on the quality of results using this method.

\begin{figure} [h!]
    \centering
    \includegraphics[width =\linewidth]{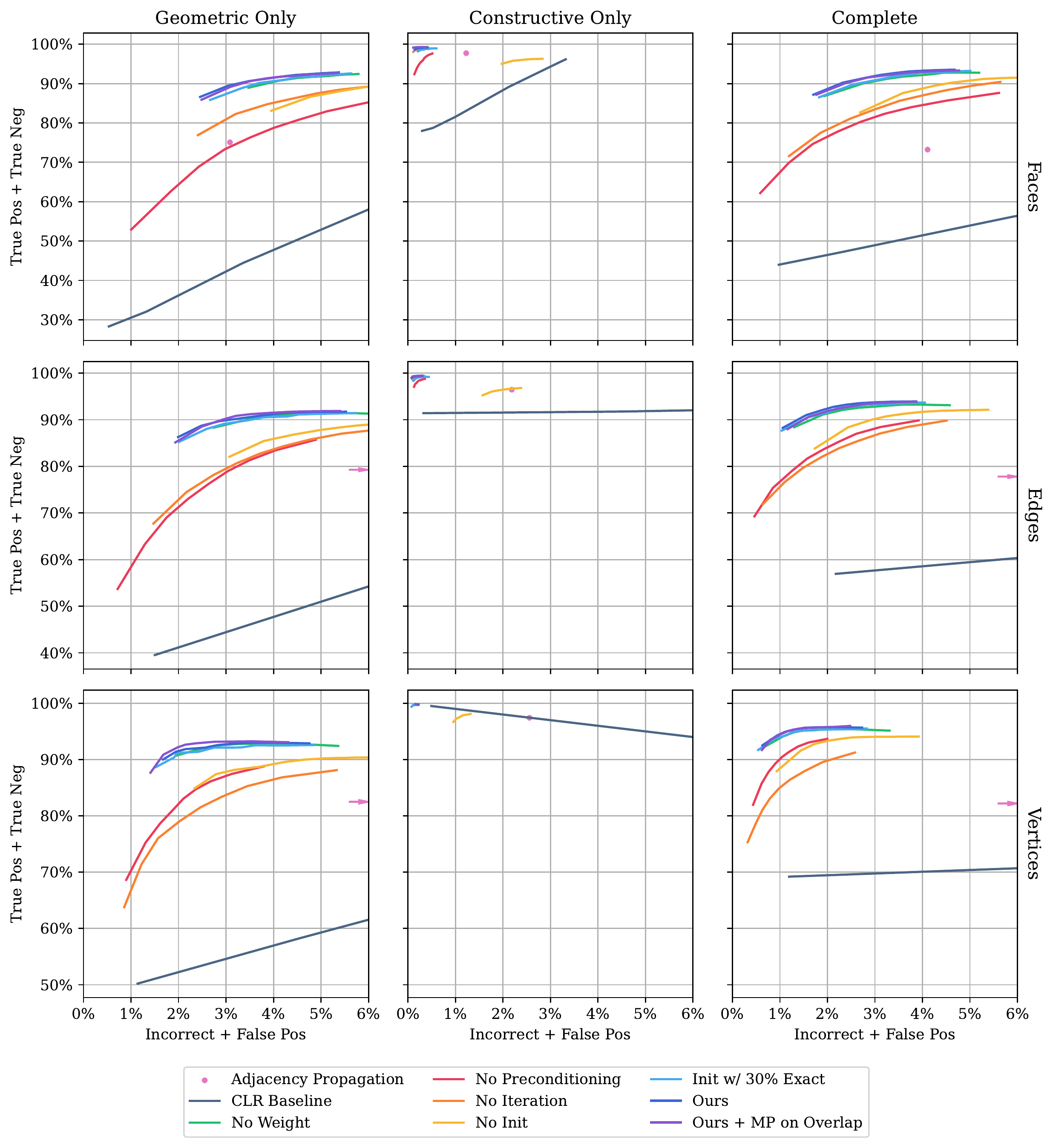}
    \caption{Comparisons between variations of our method. The plot shows trade-offs between correct labels (true positive and true negative) and incorrect labels (incorrect and false positive) (upper-left is better) The adjacency propagation baseline is shown as a single point for reference.}
    \label{fig:ablationsPareto}
\end{figure}

\paragraph{Ablations}
We compare the performance of our method against the adjacency propagation baseline, the presence or absence of various architectural choices, and the CLR baseline, and we show dominance of our final method using Pareto plots in Figure~\ref{fig:ablationsPareto}. This plot reports the possible trade-offs between correct (true positive and true negative) and incorrect (incorrect match and false positive) labels. Our method significantly outperforms the adjacency propagation baseline (represented by the pink dot in the plots), especially in terms of the amount of false matches found. 

Other design choices we test are preconditioning on partially known and incrementally inferred matches, weighted binary cross-entropy loss, and graph message passing layers along overlap matches (adding them to the match condition GAT as an additional edge type). We note that the iterative approach and preconditioning on partial matches are the most important components we evaluated, as removing them leads to a significant worsening of results across all combinations of topology types and datasets (see the red and orange curves). Both of these curves represent the performance of the single-shot algorithm where all scores are predicted at once, one with preconditioning on exact partial matches and one without. We also show that removing the weighted cross entropy loss that weights incorrect matches more heavily (green) harms performance, especially increasing the incorrect matches over the other methods compared. Finally, we evaluate the effect of additional message passing layers between entities matched by overlap (purple), but this did not consistently improve results to warrant the increased complexity, so it was excluded from our final method (blue).

We also evaluate the sensitivity from initialization by initializing our method with different percentages of exact matches. Initializing with only 30\% of the exact matches reduces performance by about 1 percentage points (cyan). However, removing initialization completely reduces performance by 8.1, 7.2, and 5.7 percentage points for faces, edges, and vertices, respectively (yellow).

\begin{figure*}[h!]
    \centering
    \includegraphics[width= \linewidth]{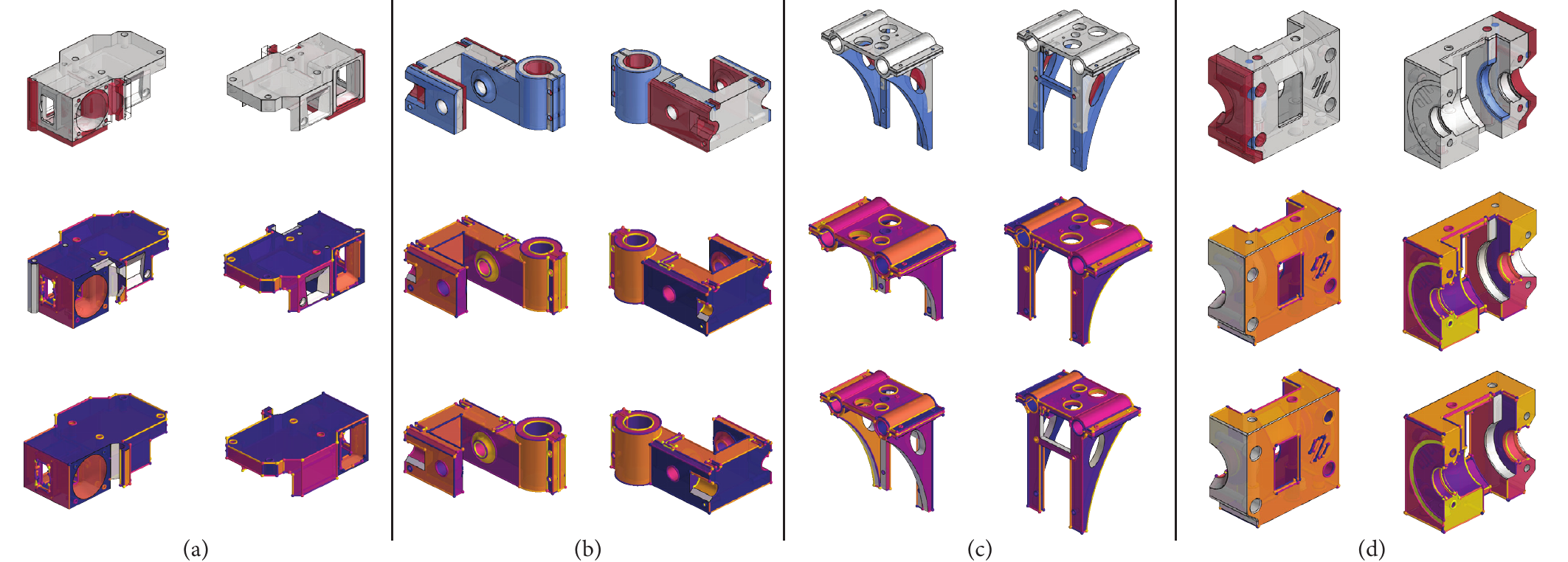}
    \caption{Examples of models from our expert data set where matches are shown from two views (columns). The top row shows shows the difference between original and updated B-reps. The second and third rows show the original and updated B-reps, respectively, with faces color-coded based on matches found by our algorithm (no ground truth is reported).   }
    \label{fig:expertViz}
\end{figure*}

\begin{figure}[h!]
    \centering
    \includegraphics[width =\linewidth]{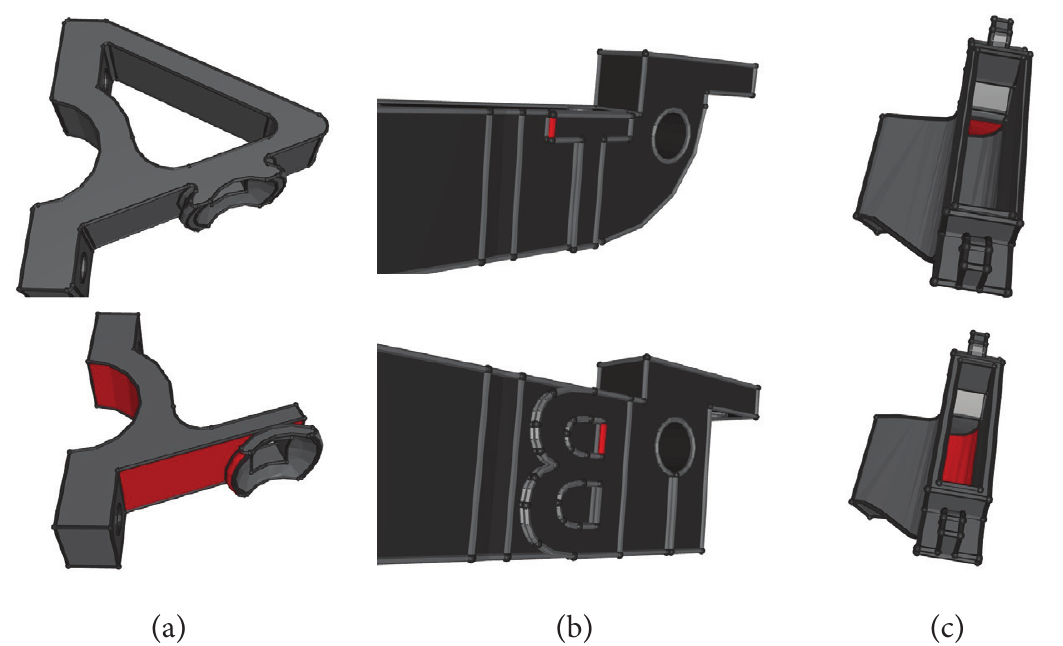}
    \caption{Examples of failure cases on the expert dataset. Some of the cases where this happens are large geometric changes (a), total replacement of sections of a part (b), and topological changes that confuse neighboring matches (c).}
    \label{fig:failedCases}
\end{figure}

\subsection{Expert Data Evaluation}

We evaluate our method using the expert collected dataset described in Section~\ref{sec:datacollection}. Because this dataset does not contain ground truth labels, we showed our method's predictions to a CAD expert and asked them to annotate matches they perceived as missing, incorrect, or extraneous through the use of an interactive GUI. The interface presented them with paired 3D visualizations with our predicted matches highlighted on a 3D model of both the original part and the variation 
and asked them whether the corresponding matches (or lack thereof) contained any errors. When a match was found, they might report that either no match should have been found, or the match found was incorrect. In addition, we filtered the expert collection to 16 examples which had a reasonable number of entities to manually annotate.

Statistics over this dataset are reported in Table \ref{tab:expert} validating whether our method trained on synthetic data generalizes to real-word examples created by CAD designers. We illustrate some of these matches in Figure~\ref{fig:expertViz} which show a large number of accurate matches for substantial geometric and topological variations.

Examples of errors found by the expert are shown in Figure~\ref{fig:failedCases}. Missed matches (a) are the most common type of error, owing to the difficulty of tracking the correspondence of topological entities that have changed geometrically (which can be an ambiguous task). In rarer cases, structural changes can lead to topological entities neighboring the change being mismatched (Figure~\ref{fig:failedCases}~(b)). Finally, the rarest type of error (occurring in less than 1\% of entities) is for the model to predict a match where none should exist. For instance, when parts of the model drastically change, such as the ``T'' shape turning into a ``B'' in Figure~\ref{fig:failedCases}~(c), our model may find some spurious matches along the letter profiles, despite the different-shaped letters not corresponding in any meaningful way. 

\begin{table}[h!]
\small
\caption{Our results on the expert collection validated with user study. All numbers are percentages of $\mathcal{B}_u$ entities.}
\begin{tabular}{lrrrr}
\toprule
& & \multicolumn{1}{c}{Faces} & \multicolumn{1}{c}{Edges} & \multicolumn{1}{c}{Vertices} \\
\midrule
Exact & True Positive & 32.1\% & 40.9\% & 45.3\% \\
\cmidrule(lr){1-5}
\textbf{Ours} & True Positive & 74.2\% & 65.1\% & 61.0\% \\
& True Negative & 15.3\% & 22.0\% & 27.8\% \\
& Missed & 8.2\% & 11.1\% & 9.9\% \\
& Incorrect & 1.8\% & 1.7\% & 0.5\% \\
& False Positive & 0.5\% & 0.0\% & 0.9\% \\
\cmidrule(r){2-5}
& Correct Label & 89.5\% & 87.1\% & 88.8\% \\
& Incorrect Label & 2.3\% & 1.7\% & 1.4\% \\
\bottomrule
\end{tabular}
\label{tab:expert}
\end{table}

\subsection{Limitations and Future Work}

A fundamental limitation of our method is that correspondences are often ambiguous and users will disagree about how matches propagate with changes. While we have used Onshape's programmatic tracking scheme as ground truth, that algorithm has its own limitations as it is driven by heuristics and may not match users' expectations in every circumstance. We argue, however, that while such ambiguities are common and will be present in a large portion of models, they typically account for a small portion of the entities within these models. This is what ensures the high performance of our method: typical models have hundreds of entities and only a very small portion of them will have ambiguous matches.  

\begin{figure} [h!]
    \centering
    \includegraphics[width = \linewidth]{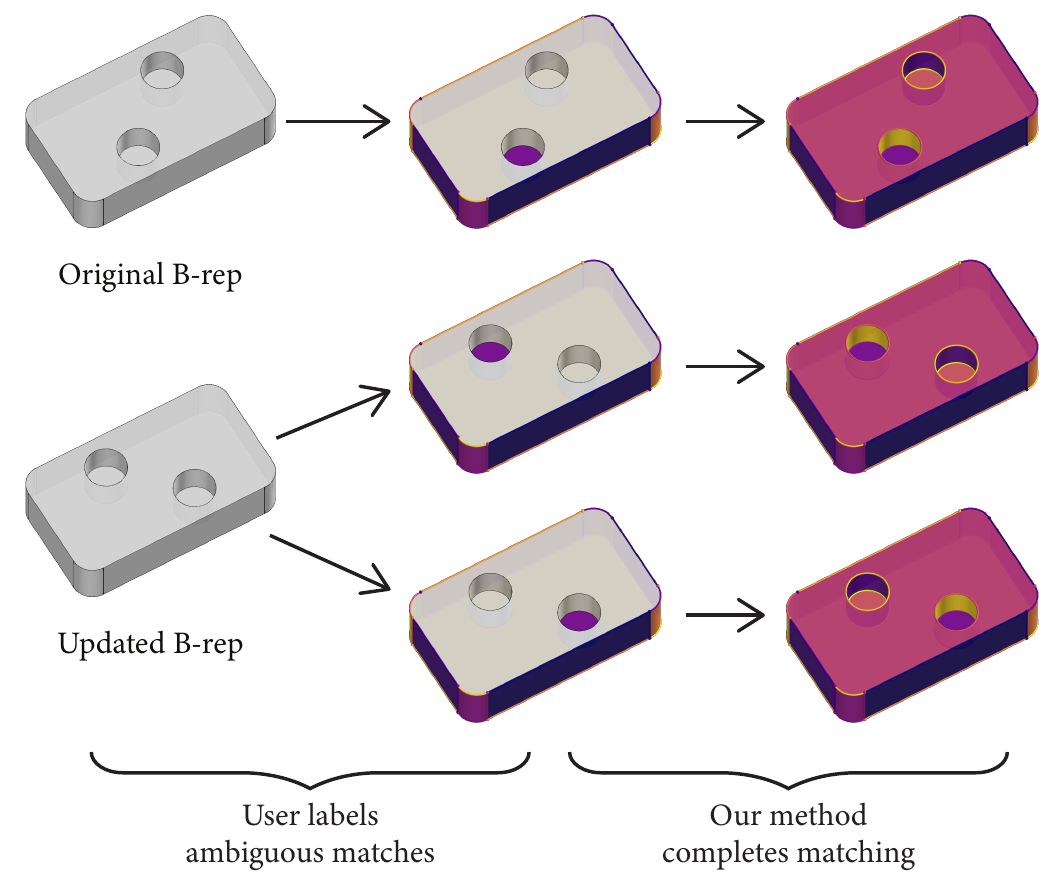}
    \caption{An example of user guidance that can be further explored in future work. Here the updated B-rep includes a 90-degree rotation of the two circular cutouts making the match ambiguous. With only \emph{one} user-specified match (face in purple), our algorithm is able to infer the rest of the matched entities.  }
    \label{fig:userGuidance}
\end{figure}

One way to address this ambiguity in future work is to incorporate a user in the loop. Since our method is iterative and driven by previously known matches, it lends itself well to this type of workflow. For example, future directions could explore additional message passes across user-specified matches for resolving ambiguities, as prototyped in Figure~\ref{fig:userGuidance}. Efforts to address ambiguities should also consider one-to-many and many-to-one matches. CAD referencing schemes that have access to the program history can output one-to-many or many-to-one correspondences. Since finding such maps is challenging with no contextual information and can lead to errors, we have chosen a conservative approach that only maps one entity to another, randomly selecting a match as ground truth if the historical referencing scheme returns a one-to-many match. Future directions should consider both one-to-many and many-to-one matches.

Another consideration is that many CAD models are made of multiple bodies and those can be exported in a single B-rep file. Out of the 25 expert examples we collected, 8 of them had more than one body as part of the original model. Since the algorithm we discussed in this paper is focused on single-body matching, for the experiments we reported above, we used only a single body as input by either selecting the main body or unioning bodies together. One option to handle multiple bodies is to first match the bodies and then use our method on each body. While body-matching can be addressed by user input or using unique names (body names are exported in the standard B-rep format), automatic body-matching is a potential area of future research. Another approach is to use our method directly on the entities of all bodies. While there are no representation constraints that would inhibit our architecture from handling this type of data, future work should collect a dataset of multiple bodies for further training and validation.

Lastly, there are many additional geometric features that we might consider as additional inputs to our system, potentially strengthening results. In particular, shape descriptors with proven utility in the related area of shape matching might be employed~\cite{gal2006salient}, as well as the geometric and topological descriptors used in prior CAD model retrieval literature~\cite{tao2013partial}. These methods could also be employed as stronger baselines---our current baselines depend on exactly overlapping geometry, and fail to capture common rotations or translations of pieces. We leave it to future work to adapt these approaches to the setting of producing conservative matches between discrete B-rep elements.

\section{Conclusion}

In this work, we present the first fully-automatic learning-based approach for matching topological entities across variations of CAD models and show that it outperforms existing baselines by thoroughly evaluating it against synthetic and expert data. We motivate the need for such an algorithm as an essential part of the CAD pipeline to enable collaborative workflows involving model re-imports. In addition to directly addressing this need, our proposed method and accompanying dataset pave the way for new research opportunities. 

The first opportunity lies within the space of CAD manipulation. Direct editing systems have been proposed to make CAD manipulation easier, but they lack the generality and consistency that can be achieved through program representations. While some works address the problem of combining direct manipulation and program representations, they struggle with persistent referencing when direct editing is used and the program representation cannot be kept in sync~\cite{Chugh:2016:1,hempel2019sketch,cascaval2022bidirectional,MB:2021:DAGA}. By enabling references to be propagated directly over geometry, the methods proposed in this paper can potentially enable novel CAD manipulation workflows.  

The second opportunity lies in transferring this information to other geometric formats, such as meshes. While shape correspondence on geometry represented as meshes or point clouds is a well-studied problem, transferring shape correspondence learned from CAD B-reps to these domains may provide a fresh perspective to the problem of shape correspondence on CAD-adjacent data, such as the vast amount of CAD models that are only available in mesh form.

\section{Acknowledgements}
This work was supported by NSF awards 2219864, 2212049, and 2017927 as well as gifts from Adobe, Intel, Meta, and Amazon. The authors further thank the whole Onshape, PTC team for their guidance and technical support. James Noeckel also  acknowledges the funding he received from UW Reality Lab, Meta, Google, OPPO, and Amazon.

\bibliographystyle{ACM-Reference-Format}
\bibliography{brepMatching,cad,schulz-edited}


\begin{thebibliography}{38}


\ifx \showCODEN    \undefined \def \showCODEN     #1{\unskip}     \fi
\ifx \showDOI      \undefined \def \showDOI       #1{#1}\fi
\ifx \showISBNx    \undefined \def \showISBNx     #1{\unskip}     \fi
\ifx \showISBNxiii \undefined \def \showISBNxiii  #1{\unskip}     \fi
\ifx \showISSN     \undefined \def \showISSN      #1{\unskip}     \fi
\ifx \showLCCN     \undefined \def \showLCCN      #1{\unskip}     \fi
\ifx \shownote     \undefined \def \shownote      #1{#1}          \fi
\ifx \showarticletitle \undefined \def \showarticletitle #1{#1}   \fi
\ifx \showURL      \undefined \def \showURL       {\relax}        \fi
\providecommand\bibfield[2]{#2}
\providecommand\bibinfo[2]{#2}
\providecommand\natexlab[1]{#1}
\providecommand\showeprint[2][]{arXiv:#2}

\bibitem[\protect\citeauthoryear{Bian, Grandi, Hassani, Sadler, Borijin,
  Fernandes, Wang, Lu, Otis, Ho, and Li}{shi}{2022}]%
        {shijie2022material}
 \bibinfo{year}{2022}\natexlab{}.
\newblock \bibinfo{booktitle}{\emph{{Material Prediction for Design Automation
  Using Graph Representation Learning}}}. \bibinfo{series}{International Design
  Engineering Technical Conferences and Computers and Information in
  Engineering Conference}, Vol.~\bibinfo{volume}{Volume 3A: 48th Design
  Automation Conference (DAC)}.
\newblock
\urldef\tempurl%
\url{https://doi.org/10.1115/DETC2022-88049}
\showDOI{\tempurl}
\showeprint{https://asmedigitalcollection.asme.org/IDETC-CIE/proceedings-pdf/IDETC-CIE2022/86229/V03AT03A001/6943080/v03at03a001-detc2022-88049.pdf}
\newblock
\shownote{V03AT03A001.}


\bibitem[\protect\citeauthoryear{Baran and Schulz}{Baran and Schulz}{2020}]%
        {baran:2020:brepmatching}
\bibfield{author}{\bibinfo{person}{Ilya Baran} {and} \bibinfo{person}{Adriana
  Schulz}.} \bibinfo{year}{2020}\natexlab{}.
\newblock \bibinfo{title}{B-rep matching for maintaining associativity across
  CAD interoperation}.
\newblock
\newblock
\newblock
\shownote{US Patent App. 16/735,194.}


\bibitem[\protect\citeauthoryear{Bharadwaj, Xu, Angrish, Chen, and
  Starly}{Bharadwaj et~al\mbox{.}}{2019}]%
        {starly2019fabwave}
\bibfield{author}{\bibinfo{person}{Akshay Bharadwaj}, \bibinfo{person}{Yang
  Xu}, \bibinfo{person}{Atin Angrish}, \bibinfo{person}{Yong Chen}, {and}
  \bibinfo{person}{Binil Starly}.} \bibinfo{year}{2019}\natexlab{}.
\newblock \showarticletitle{{Development of a Pilot Manufacturing
  Cyberinfrastructure With an Information Rich Mechanical CAD 3D Model
  Repository}} \emph{(\bibinfo{series}{International Manufacturing Science and
  Engineering Conference}, Vol.~\bibinfo{volume}{Volume 1: Additive
  Manufacturing; Manufacturing Equipment and Systems; Bio and Sustainable
  Manufacturing})}.
\newblock
\urldef\tempurl%
\url{https://doi.org/10.1115/MSEC2019-2882}
\showDOI{\tempurl}


\bibitem[\protect\citeauthoryear{Bidarra and Bronsvoort}{Bidarra and
  Bronsvoort}{2000}]%
        {bidarra2000semantic}
\bibfield{author}{\bibinfo{person}{Rafael Bidarra} {and}
  \bibinfo{person}{Willem~F Bronsvoort}.} \bibinfo{year}{2000}\natexlab{}.
\newblock \showarticletitle{Semantic feature modelling}.
\newblock \bibinfo{journal}{\emph{Computer-Aided Design}} \bibinfo{volume}{32},
  \bibinfo{number}{3} (\bibinfo{year}{2000}), \bibinfo{pages}{201--225}.
\newblock


\bibitem[\protect\citeauthoryear{Bidarra, Nyirenda, and Bronsvoort}{Bidarra
  et~al\mbox{.}}{2005}]%
        {bidarra2005feature}
\bibfield{author}{\bibinfo{person}{Rafael Bidarra}, \bibinfo{person}{Paulos~J
  Nyirenda}, {and} \bibinfo{person}{Willem~F Bronsvoort}.}
  \bibinfo{year}{2005}\natexlab{}.
\newblock \showarticletitle{A feature-based solution to the persistent naming
  problem}.
\newblock \bibinfo{journal}{\emph{Computer-Aided Design and Applications}}
  \bibinfo{volume}{2}, \bibinfo{number}{1-4} (\bibinfo{year}{2005}),
  \bibinfo{pages}{517--526}.
\newblock


\bibitem[\protect\citeauthoryear{Cao, Robinson, Hua, Boussuge, Colligan, and
  Pan}{Cao et~al\mbox{.}}{2020}]%
        {cao2020graph}
\bibfield{author}{\bibinfo{person}{Weijuan Cao}, \bibinfo{person}{Trevor
  Robinson}, \bibinfo{person}{Yang Hua}, \bibinfo{person}{Flavien Boussuge},
  \bibinfo{person}{Andrew~R. Colligan}, {and} \bibinfo{person}{Wanbin Pan}.}
  \bibinfo{year}{2020}\natexlab{}.
\newblock \showarticletitle{Graph Representation of 3D CAD Models for Machining
  Feature Recognition With Deep Learning} \emph{(\bibinfo{series}{International
  Design Engineering Technical Conferences and Computers and Information in
  Engineering Conference}, Vol.~\bibinfo{volume}{Volume 11A: 46th Design
  Automation Conference (DAC)})}.
\newblock


\bibitem[\protect\citeauthoryear{Cascaval, Shalah, Quinn, Bod{\'{\i}}k,
  Agrawala, and Schulz}{Cascaval et~al\mbox{.}}{2021}]%
        {cascaval2022bidirectional}
\bibfield{author}{\bibinfo{person}{Dan Cascaval}, \bibinfo{person}{Mira
  Shalah}, \bibinfo{person}{Phillip Quinn}, \bibinfo{person}{Rastislav
  Bod{\'{\i}}k}, \bibinfo{person}{Maneesh Agrawala}, {and}
  \bibinfo{person}{Adriana Schulz}.} \bibinfo{year}{2021}\natexlab{}.
\newblock \showarticletitle{Differentiable 3D {CAD} Programs for Bidirectional
  Editing}.
\newblock \bibinfo{journal}{\emph{CoRR}}  \bibinfo{volume}{abs/2110.01182}
  (\bibinfo{year}{2021}).
\newblock
\showeprint[arXiv]{2110.01182}
\urldef\tempurl%
\url{https://arxiv.org/abs/2110.01182}
\showURL{%
\tempurl}


\bibitem[\protect\citeauthoryear{Cheon, Mun, Han, and Kim}{Cheon
  et~al\mbox{.}}{2012}]%
        {cheon2012name}
\bibfield{author}{\bibinfo{person}{Sang-Uk Cheon}, \bibinfo{person}{Duhwan
  Mun}, \bibinfo{person}{Soonhung Han}, {and} \bibinfo{person}{Byung~Chul
  Kim}.} \bibinfo{year}{2012}\natexlab{}.
\newblock \showarticletitle{Name matching method using topology merging and
  splitting history for exchange of feature-based CAD models}.
\newblock \bibinfo{journal}{\emph{Journal of mechanical science and
  technology}} \bibinfo{volume}{26}, \bibinfo{number}{10}
  (\bibinfo{year}{2012}), \bibinfo{pages}{3201--3212}.
\newblock


\bibitem[\protect\citeauthoryear{Chugh, Hempel, Spradlin, and Albers}{Chugh
  et~al\mbox{.}}{2016}]%
        {Chugh:2016:1}
\bibfield{author}{\bibinfo{person}{Ravi Chugh}, \bibinfo{person}{Brian Hempel},
  \bibinfo{person}{Mitchell Spradlin}, {and} \bibinfo{person}{Jacob Albers}.}
  \bibinfo{year}{2016}\natexlab{}.
\newblock \showarticletitle{Programmatic and Direct Manipulation, Together at
  Last}. In \bibinfo{booktitle}{\emph{Proceedings of the 37th ACM SIGPLAN
  Conference on Programming Language Design and Implementation}}
  \emph{(\bibinfo{series}{PLDI '16})}. ACM, \bibinfo{address}{Santa Barbara,
  CA, USA}, \bibinfo{pages}{341--354}.
\newblock


\bibitem[\protect\citeauthoryear{Deng, Yao, Dyke, and Zhang}{Deng
  et~al\mbox{.}}{2022}]%
        {deng2022survey}
\bibfield{author}{\bibinfo{person}{Bailin Deng}, \bibinfo{person}{Yuxin Yao},
  \bibinfo{person}{Roberto~M Dyke}, {and} \bibinfo{person}{Juyong Zhang}.}
  \bibinfo{year}{2022}\natexlab{}.
\newblock \showarticletitle{A Survey of Non-Rigid 3D Registration}. In
  \bibinfo{booktitle}{\emph{Computer Graphics Forum}},
  Vol.~\bibinfo{volume}{41}. Wiley Online Library, \bibinfo{pages}{559--589}.
\newblock


\bibitem[\protect\citeauthoryear{Dorribo~Camba, Contero,
  et~al\mbox{.}}{Dorribo~Camba et~al\mbox{.}}{2016}]%
        {dorribo2016parametric}
\bibfield{author}{\bibinfo{person}{Jorge Dorribo~Camba},
  \bibinfo{person}{Manuel Contero}, {et~al\mbox{.}}}
  \bibinfo{year}{2016}\natexlab{}.
\newblock \showarticletitle{Parametric CAD modeling: An analysis of strategies
  for design reusability}.
\newblock  (\bibinfo{year}{2016}).
\newblock


\bibitem[\protect\citeauthoryear{Farjana and Han}{Farjana and Han}{2018}]%
        {farjana2018mechanisms}
\bibfield{author}{\bibinfo{person}{Shahjadi~Hisan Farjana} {and}
  \bibinfo{person}{Soonhung Han}.} \bibinfo{year}{2018}\natexlab{}.
\newblock \showarticletitle{Mechanisms of persistent identification of
  topological entities in CAD systems: A review}.
\newblock \bibinfo{journal}{\emph{Alexandria engineering journal}}
  \bibinfo{volume}{57}, \bibinfo{number}{4} (\bibinfo{year}{2018}),
  \bibinfo{pages}{2837--2849}.
\newblock


\bibitem[\protect\citeauthoryear{Gal and Cohen-Or}{Gal and Cohen-Or}{2006}]%
        {gal2006salient}
\bibfield{author}{\bibinfo{person}{Ran Gal} {and} \bibinfo{person}{Daniel
  Cohen-Or}.} \bibinfo{year}{2006}\natexlab{}.
\newblock \showarticletitle{Salient geometric features for partial shape
  matching and similarity}.
\newblock \bibinfo{journal}{\emph{ACM Transactions on Graphics}}
  \bibinfo{volume}{25}, \bibinfo{number}{1} (\bibinfo{date}{Jan.}
  \bibinfo{year}{2006}), \bibinfo{pages}{130--150}.
\newblock
\showISSN{0730-0301, 1557-7368}
\urldef\tempurl%
\url{https://doi.org/10.1145/1122501.1122507}
\showDOI{\tempurl}


\bibitem[\protect\citeauthoryear{Guo, Liu, Pan, Yang, Tong, and Guo}{Guo
  et~al\mbox{.}}{2022}]%
        {Guo2022Complexgen}
\bibfield{author}{\bibinfo{person}{Hao-Xiang Guo}, \bibinfo{person}{Shilin
  Liu}, \bibinfo{person}{Hao Pan}, \bibinfo{person}{Liu Yang},
  \bibinfo{person}{Xin Tong}, {and} \bibinfo{person}{Baining Guo}.}
  \bibinfo{year}{2022}\natexlab{}.
\newblock \showarticletitle{ComplexGen: CAD Reconstruction by B-Rep Chain
  Complex Generation}.
\newblock \bibinfo{journal}{\emph{ACM Transactions on Graphics (TOG)}}
  \bibinfo{volume}{39}, \bibinfo{number}{4} (\bibinfo{year}{2022}),
  \bibinfo{pages}{106:1--106:14}.
\newblock


\bibitem[\protect\citeauthoryear{Hempel, Lubin, and Chugh}{Hempel
  et~al\mbox{.}}{2019}]%
        {hempel2019sketch}
\bibfield{author}{\bibinfo{person}{Brian Hempel}, \bibinfo{person}{Justin
  Lubin}, {and} \bibinfo{person}{Ravi Chugh}.} \bibinfo{year}{2019}\natexlab{}.
\newblock \showarticletitle{Sketch-n-Sketch: Output-Directed Programming for
  SVG}. In \bibinfo{booktitle}{\emph{Proceedings of the 32nd Annual ACM
  Symposium on User Interface Software and Technology}}.
  \bibinfo{pages}{281--292}.
\newblock


\bibitem[\protect\citeauthoryear{Jayaraman, Lambourne, Desai, Willis, Sanghi,
  and Morris}{Jayaraman et~al\mbox{.}}{2022}]%
        {jayaraman2022solidgen}
\bibfield{author}{\bibinfo{person}{Pradeep~Kumar Jayaraman},
  \bibinfo{person}{Joseph~G Lambourne}, \bibinfo{person}{Nishkrit Desai},
  \bibinfo{person}{Karl~DD Willis}, \bibinfo{person}{Aditya Sanghi}, {and}
  \bibinfo{person}{Nigel~JW Morris}.} \bibinfo{year}{2022}\natexlab{}.
\newblock \showarticletitle{SolidGen: An Autoregressive Model for Direct B-rep
  Synthesis}.
\newblock \bibinfo{journal}{\emph{arXiv preprint arXiv:2203.13944}}
  (\bibinfo{year}{2022}).
\newblock


\bibitem[\protect\citeauthoryear{Jayaraman, Sanghi, Lambourne, Willis, Davies,
  Shayani, and Morris}{Jayaraman et~al\mbox{.}}{2021}]%
        {jayaraman2021uv}
\bibfield{author}{\bibinfo{person}{Pradeep~Kumar Jayaraman},
  \bibinfo{person}{Aditya Sanghi}, \bibinfo{person}{Joseph~G Lambourne},
  \bibinfo{person}{Karl~DD Willis}, \bibinfo{person}{Thomas Davies},
  \bibinfo{person}{Hooman Shayani}, {and} \bibinfo{person}{Nigel Morris}.}
  \bibinfo{year}{2021}\natexlab{}.
\newblock \showarticletitle{UV-Net: Learning From Boundary Representations}. In
  \bibinfo{booktitle}{\emph{Proceedings of the IEEE/CVF Conference on Computer
  Vision and Pattern Recognition}}. \bibinfo{pages}{11703--11712}.
\newblock


\bibitem[\protect\citeauthoryear{Jones, Hildreth, Chen, Baran, Kim, and
  Schulz}{Jones et~al\mbox{.}}{2021}]%
        {jones:2021:automate}
\bibfield{author}{\bibinfo{person}{Benjamin Jones}, \bibinfo{person}{Dalton
  Hildreth}, \bibinfo{person}{Duowen Chen}, \bibinfo{person}{Ilya Baran},
  \bibinfo{person}{Vladimir~G. Kim}, {and} \bibinfo{person}{Adriana Schulz}.}
  \bibinfo{year}{2021}\natexlab{}.
\newblock \showarticletitle{AutoMate: A Dataset and Learning Approach for
  Automatic Mating of CAD Assemblies}.
\newblock \bibinfo{journal}{\emph{ACM Transactions on Graphics}}
  \bibinfo{volume}{40}, \bibinfo{number}{6}, Article \bibinfo{articleno}{227}
  (\bibinfo{date}{dec} \bibinfo{year}{2021}), \bibinfo{numpages}{18}~pages.
\newblock
\showISSN{0730-0301}
\urldef\tempurl%
\url{https://doi.org/10.1145/3478513.3480562}
\showDOI{\tempurl}


\bibitem[\protect\citeauthoryear{Jones, Hu, Kim, and Schulz}{Jones
  et~al\mbox{.}}{2022}]%
        {jones2022self}
\bibfield{author}{\bibinfo{person}{Benjamin~T. Jones}, \bibinfo{person}{Michael
  Hu}, \bibinfo{person}{Vladimir~G. Kim}, {and} \bibinfo{person}{Adriana
  Schulz}.} \bibinfo{year}{2022}\natexlab{}.
\newblock \bibinfo{title}{Self-Supervised Representation Learning for CAD}.
\newblock
\newblock
\urldef\tempurl%
\url{https://doi.org/10.48550/ARXIV.2210.10807}
\showDOI{\tempurl}


\bibitem[\protect\citeauthoryear{Kirkwood and Sherwood}{Kirkwood and
  Sherwood}{2018}]%
        {kirkwood2018sustained}
\bibfield{author}{\bibinfo{person}{Robert Kirkwood} {and}
  \bibinfo{person}{James~A Sherwood}.} \bibinfo{year}{2018}\natexlab{}.
\newblock \showarticletitle{Sustained CAD/CAE integration: integrating with
  successive versions of step or IGES files}.
\newblock \bibinfo{journal}{\emph{Engineering with Computers}}
  \bibinfo{volume}{34}, \bibinfo{number}{1} (\bibinfo{year}{2018}),
  \bibinfo{pages}{1--13}.
\newblock


\bibitem[\protect\citeauthoryear{Koch, Matveev, Jiang, Williams, Artemov,
  Burnaev, Alexa, Zorin, and Panozzo}{Koch et~al\mbox{.}}{2019}]%
        {koch_abc_2019}
\bibfield{author}{\bibinfo{person}{Sebastian Koch}, \bibinfo{person}{Albert
  Matveev}, \bibinfo{person}{Zhongshi Jiang}, \bibinfo{person}{Francis
  Williams}, \bibinfo{person}{Alexey Artemov}, \bibinfo{person}{Evgeny
  Burnaev}, \bibinfo{person}{Marc Alexa}, \bibinfo{person}{Denis Zorin}, {and}
  \bibinfo{person}{Daniele Panozzo}.} \bibinfo{year}{2019}\natexlab{}.
\newblock \showarticletitle{{ABC}: {A} {Big} {CAD} {Model} {Dataset} for
  {Geometric} {Deep} {Learning}}. In \bibinfo{booktitle}{\emph{2019
  {IEEE}/{CVF} {Conference} on {Computer} {Vision} and {Pattern} {Recognition}
  ({CVPR})}}. \bibinfo{publisher}{IEEE}, \bibinfo{address}{Long Beach, CA,
  USA}, \bibinfo{pages}{9593--9603}.
\newblock
\showISBNx{978-1-72813-293-8}
\urldef\tempurl%
\url{https://doi.org/10.1109/CVPR.2019.00983}
\showDOI{\tempurl}


\bibitem[\protect\citeauthoryear{Lambourne, Willis, Jayaraman, Zhang, Sanghi,
  and Malekshan}{Lambourne et~al\mbox{.}}{2022}]%
        {rahimi2022reconstructing}
\bibfield{author}{\bibinfo{person}{Joseph~George Lambourne},
  \bibinfo{person}{Karl Willis}, \bibinfo{person}{Pradeep~Kumar Jayaraman},
  \bibinfo{person}{Longfei Zhang}, \bibinfo{person}{Aditya Sanghi}, {and}
  \bibinfo{person}{Kamal~Rahimi Malekshan}.} \bibinfo{year}{2022}\natexlab{}.
\newblock \showarticletitle{Reconstructing Editable Prismatic CAD from Rounded
  Voxel Models}. In \bibinfo{booktitle}{\emph{SIGGRAPH Asia 2022 Conference
  Papers}} (Daegu, Republic of Korea) \emph{(\bibinfo{series}{SA '22})}.
  \bibinfo{publisher}{Association for Computing Machinery},
  \bibinfo{address}{New York, NY, USA}, Article \bibinfo{articleno}{53},
  \bibinfo{numpages}{9}~pages.
\newblock
\showISBNx{9781450394703}
\urldef\tempurl%
\url{https://doi.org/10.1145/3550469.3555424}
\showDOI{\tempurl}


\bibitem[\protect\citeauthoryear{Lambourne, Willis, Jayaraman, Sanghi, Meltzer,
  and Shayani}{Lambourne et~al\mbox{.}}{2021}]%
        {lambourne_brepnet_2021}
\bibfield{author}{\bibinfo{person}{Joseph~G. Lambourne}, \bibinfo{person}{Karl
  D.~D. Willis}, \bibinfo{person}{Pradeep~Kumar Jayaraman},
  \bibinfo{person}{Aditya Sanghi}, \bibinfo{person}{Peter Meltzer}, {and}
  \bibinfo{person}{Hooman Shayani}.} \bibinfo{year}{2021}\natexlab{}.
\newblock \showarticletitle{{BRepNet}: {A} topological message passing system
  for solid models}.
\newblock \bibinfo{journal}{\emph{arXiv:2104.00706 [cs]}}
  (\bibinfo{date}{April} \bibinfo{year}{2021}).
\newblock
\urldef\tempurl%
\url{http://arxiv.org/abs/2104.00706}
\showURL{%
\tempurl}
\newblock
\shownote{arXiv: 2104.00706.}


\bibitem[\protect\citeauthoryear{Lupinetti, Pernot, Monti, and
  Giannini}{Lupinetti et~al\mbox{.}}{2019}]%
        {lupinetti2019content}
\bibfield{author}{\bibinfo{person}{Katia Lupinetti},
  \bibinfo{person}{Jean-Philippe Pernot}, \bibinfo{person}{Marina Monti}, {and}
  \bibinfo{person}{Franca Giannini}.} \bibinfo{year}{2019}\natexlab{}.
\newblock \showarticletitle{Content-based {CAD} assembly model retrieval:
  {Survey} and future challenges}.
\newblock \bibinfo{journal}{\emph{Computer-Aided Design}}
  \bibinfo{volume}{113}, \bibinfo{number}{C} (\bibinfo{date}{Aug.}
  \bibinfo{year}{2019}), \bibinfo{pages}{62--81}.
\newblock
\showISSN{0010-4485}
\urldef\tempurl%
\url{https://doi.org/10.1016/j.cad.2019.03.005}
\showDOI{\tempurl}


\bibitem[\protect\citeauthoryear{Michel and Boubekeur}{Michel and
  Boubekeur}{2021}]%
        {MB:2021:DAGA}
\bibfield{author}{\bibinfo{person}{Elie Michel} {and} \bibinfo{person}{Tamy
  Boubekeur}.} \bibinfo{year}{2021}\natexlab{}.
\newblock \showarticletitle{DAG Amendment for Inverse Control of Parametric
  Shapes}.
\newblock \bibinfo{journal}{\emph{ACM Transactions on Graphics}}
  \bibinfo{volume}{40}, \bibinfo{number}{4} (\bibinfo{year}{2021}),
  \bibinfo{pages}{173:1--173:14}.
\newblock


\bibitem[\protect\citeauthoryear{Sahillio{\u{g}}lu}{Sahillio{\u{g}}lu}{2020}]%
        {sahilliouglu2020recent}
\bibfield{author}{\bibinfo{person}{Yusuf Sahillio{\u{g}}lu}.}
  \bibinfo{year}{2020}\natexlab{}.
\newblock \showarticletitle{Recent advances in shape correspondence}.
\newblock \bibinfo{journal}{\emph{The Visual Computer}} \bibinfo{volume}{36},
  \bibinfo{number}{8} (\bibinfo{year}{2020}), \bibinfo{pages}{1705--1721}.
\newblock


\bibitem[\protect\citeauthoryear{Seff, Ovadia, Zhou, and Adams}{Seff
  et~al\mbox{.}}{2020}]%
        {seff2020sketchgraphs}
\bibfield{author}{\bibinfo{person}{Ari Seff}, \bibinfo{person}{Yaniv Ovadia},
  \bibinfo{person}{Wenda Zhou}, {and} \bibinfo{person}{Ryan~P Adams}.}
  \bibinfo{year}{2020}\natexlab{}.
\newblock \showarticletitle{Sketchgraphs: A large-scale dataset for modeling
  relational geometry in computer-aided design}.
\newblock \bibinfo{journal}{\emph{arXiv preprint arXiv:2007.08506}}
  (\bibinfo{year}{2020}).
\newblock


\bibitem[\protect\citeauthoryear{Spitz and Rappoport}{Spitz and
  Rappoport}{2007}]%
        {spitz2007boundary}
\bibfield{author}{\bibinfo{person}{Steven Spitz} {and} \bibinfo{person}{Ari
  Rappoport}.} \bibinfo{year}{2007}\natexlab{}.
\newblock \bibinfo{title}{Boundary representation per feature methods and
  systems}.
\newblock
\newblock
\newblock
\shownote{US Patent 7,277,835.}


\bibitem[\protect\citeauthoryear{Tao, Huang, Ma, Guo, Wang, and Xie}{Tao
  et~al\mbox{.}}{2013}]%
        {tao2013partial}
\bibfield{author}{\bibinfo{person}{Songqiao Tao}, \bibinfo{person}{Zhengdong
  Huang}, \bibinfo{person}{Lujie Ma}, \bibinfo{person}{Shunsheng Guo},
  \bibinfo{person}{Shuting Wang}, {and} \bibinfo{person}{Youbai Xie}.}
  \bibinfo{year}{2013}\natexlab{}.
\newblock \showarticletitle{Partial Retrieval of CAD Models Based on Local
  Surface Region Decomposition}.
\newblock \bibinfo{journal}{\emph{Comput. Aided Des.}} \bibinfo{volume}{45},
  \bibinfo{number}{11} (\bibinfo{date}{nov} \bibinfo{year}{2013}),
  \bibinfo{pages}{1239–1252}.
\newblock
\showISSN{0010-4485}
\urldef\tempurl%
\url{https://doi.org/10.1016/j.cad.2013.05.008}
\showDOI{\tempurl}


\bibitem[\protect\citeauthoryear{Uy, yu~Chang, Sung, Goel, Lambourne, Birdal,
  and Guibas}{Uy et~al\mbox{.}}{2022}]%
        {uy-point2cyl-cvpr22}
\bibfield{author}{\bibinfo{person}{Mikaela~Angelina Uy}, \bibinfo{person}{Yen
  yu Chang}, \bibinfo{person}{Minhyuk Sung}, \bibinfo{person}{Purvi Goel},
  \bibinfo{person}{Joseph Lambourne}, \bibinfo{person}{Tolga Birdal}, {and}
  \bibinfo{person}{Leonidas Guibas}.} \bibinfo{year}{2022}\natexlab{}.
\newblock \showarticletitle{Point2Cyl: Reverse Engineering 3D Objects from
  Point Clouds to Extrusion Cylinders}. In \bibinfo{booktitle}{\emph{Conference
  on Computer Vision and Pattern Recognition (CVPR)}}.
\newblock


\bibitem[\protect\citeauthoryear{Vandenbrande, Grandine, Lucian, and
  Monahan}{Vandenbrande et~al\mbox{.}}{2013}]%
        {vandenbrande2013methods}
\bibfield{author}{\bibinfo{person}{Jan~H Vandenbrande},
  \bibinfo{person}{Thomas~A Grandine}, \bibinfo{person}{Miriam Lucian}, {and}
  \bibinfo{person}{John Monahan}.} \bibinfo{year}{2013}\natexlab{}.
\newblock \bibinfo{title}{Methods and apparatus for automated part positioning
  based on geometrical comparisons}.
\newblock
\newblock
\newblock
\shownote{US Patent 8,576,224.}


\bibitem[\protect\citeauthoryear{Veli{\v{c}}kovi{\'{c}}, Cucurull, Casanova,
  Romero, Li{\`{o}}, and Bengio}{Veli{\v{c}}kovi{\'{c}} et~al\mbox{.}}{2018}]%
        {velickovic2018graph}
\bibfield{author}{\bibinfo{person}{Petar Veli{\v{c}}kovi{\'{c}}},
  \bibinfo{person}{Guillem Cucurull}, \bibinfo{person}{Arantxa Casanova},
  \bibinfo{person}{Adriana Romero}, \bibinfo{person}{Pietro Li{\`{o}}}, {and}
  \bibinfo{person}{Yoshua Bengio}.} \bibinfo{year}{2018}\natexlab{}.
\newblock \showarticletitle{{Graph Attention Networks}}.
\newblock \bibinfo{journal}{\emph{International Conference on Learning
  Representations}} (\bibinfo{year}{2018}).
\newblock
\urldef\tempurl%
\url{https://openreview.net/forum?id=rJXMpikCZ}
\showURL{%
\tempurl}


\bibitem[\protect\citeauthoryear{Willis, Jayaraman, Chu, Tian, Li, Grandi,
  Sanghi, Tran, Lambourne, Solar-Lezama, et~al\mbox{.}}{Willis
  et~al\mbox{.}}{2021}]%
        {willis2021joinable}
\bibfield{author}{\bibinfo{person}{Karl~DD Willis},
  \bibinfo{person}{Pradeep~Kumar Jayaraman}, \bibinfo{person}{Hang Chu},
  \bibinfo{person}{Yunsheng Tian}, \bibinfo{person}{Yifei Li},
  \bibinfo{person}{Daniele Grandi}, \bibinfo{person}{Aditya Sanghi},
  \bibinfo{person}{Linh Tran}, \bibinfo{person}{Joseph~G Lambourne},
  \bibinfo{person}{Armando Solar-Lezama}, {et~al\mbox{.}}}
  \bibinfo{year}{2021}\natexlab{}.
\newblock \showarticletitle{JoinABLe: Learning Bottom-up Assembly of Parametric
  CAD Joints}.
\newblock \bibinfo{journal}{\emph{arXiv preprint arXiv:2111.12772}}
  (\bibinfo{year}{2021}).
\newblock


\bibitem[\protect\citeauthoryear{Willis, Pu, Luo, Chu, Du, Lambourne,
  Solar-Lezama, and Matusik}{Willis et~al\mbox{.}}{2020}]%
        {willis_fusion_2020}
\bibfield{author}{\bibinfo{person}{Karl D.~D. Willis}, \bibinfo{person}{Yewen
  Pu}, \bibinfo{person}{Jieliang Luo}, \bibinfo{person}{Hang Chu},
  \bibinfo{person}{Tao Du}, \bibinfo{person}{Joseph~G. Lambourne},
  \bibinfo{person}{Armando Solar-Lezama}, {and} \bibinfo{person}{Wojciech
  Matusik}.} \bibinfo{year}{2020}\natexlab{}.
\newblock \showarticletitle{Fusion 360 {Gallery}: {A} {Dataset} and
  {Environment} for {Programmatic} {CAD} {Reconstruction}}.
\newblock \bibinfo{journal}{\emph{arXiv:2010.02392 [cs]}} (\bibinfo{date}{Oct.}
  \bibinfo{year}{2020}).
\newblock
\urldef\tempurl%
\url{http://arxiv.org/abs/2010.02392}
\showURL{%
\tempurl}
\newblock
\shownote{arXiv: 2010.02392.}


\bibitem[\protect\citeauthoryear{Wu, Xiao, and Zheng}{Wu et~al\mbox{.}}{2021}]%
        {wu_deepcad_2021}
\bibfield{author}{\bibinfo{person}{Rundi Wu}, \bibinfo{person}{Chang Xiao},
  {and} \bibinfo{person}{Changxi Zheng}.} \bibinfo{year}{2021}\natexlab{}.
\newblock \showarticletitle{{DeepCAD}: {A} {Deep} {Generative} {Network} for
  {Computer}-{Aided} {Design} {Models}}. In \bibinfo{booktitle}{\emph{2021
  {IEEE}/{CVF} {International} {Conference} on {Computer} {Vision} ({ICCV})}}.
  \bibinfo{publisher}{IEEE}, \bibinfo{address}{Montreal, QC, Canada},
  \bibinfo{pages}{6752--6762}.
\newblock
\showISBNx{978-1-66542-812-5}
\urldef\tempurl%
\url{https://doi.org/10.1109/ICCV48922.2021.00670}
\showDOI{\tempurl}


\bibitem[\protect\citeauthoryear{Xu, Peng, Cheng, Willis, and Ritchie}{Xu
  et~al\mbox{.}}{2021}]%
        {xu2021zonegraphs}
\bibfield{author}{\bibinfo{person}{Xianghao Xu}, \bibinfo{person}{Wenzhe Peng},
  \bibinfo{person}{Chin-Yi Cheng}, \bibinfo{person}{Karl D.~D. Willis}, {and}
  \bibinfo{person}{Daniel Ritchie}.} \bibinfo{year}{2021}\natexlab{}.
\newblock \showarticletitle{Inferring CAD Modeling Sequences Using Zone
  Graphs}. In \bibinfo{booktitle}{\emph{CVPR}}.
\newblock


\bibitem[\protect\citeauthoryear{Xu, Willis, Lambourne, Cheng, Jayaraman, and
  Furukawa}{Xu et~al\mbox{.}}{2022}]%
        {xu_skexgen_2022}
\bibfield{author}{\bibinfo{person}{Xiang Xu}, \bibinfo{person}{Karl~D.D.
  Willis}, \bibinfo{person}{Joseph~G Lambourne}, \bibinfo{person}{Chin-Yi
  Cheng}, \bibinfo{person}{Pradeep~Kumar Jayaraman}, {and}
  \bibinfo{person}{Yasutaka Furukawa}.} \bibinfo{year}{2022}\natexlab{}.
\newblock \showarticletitle{{S}kex{G}en: Autoregressive Generation of {CAD}
  Construction Sequences with Disentangled Codebooks}. In
  \bibinfo{booktitle}{\emph{Proceedings of the 39th International Conference on
  Machine Learning}} \emph{(\bibinfo{series}{Proceedings of Machine Learning
  Research}, Vol.~\bibinfo{volume}{162})},
  \bibfield{editor}{\bibinfo{person}{Kamalika Chaudhuri},
  \bibinfo{person}{Stefanie Jegelka}, \bibinfo{person}{Le~Song},
  \bibinfo{person}{Csaba Szepesvari}, \bibinfo{person}{Gang Niu}, {and}
  \bibinfo{person}{Sivan Sabato}} (Eds.). \bibinfo{publisher}{PMLR},
  \bibinfo{pages}{24698--24724}.
\newblock
\urldef\tempurl%
\url{https://proceedings.mlr.press/v162/xu22k.html}
\showURL{%
\tempurl}


\bibitem[\protect\citeauthoryear{Yares}{Yares}{2013}]%
        {Thefaile61:online}
\bibfield{author}{\bibinfo{person}{Evan Yares}.}
  \bibinfo{year}{2013}\natexlab{}.
\newblock \bibinfo{title}{The failed promise of parametric CAD part 1: From the
  beginning}.
\newblock
  \bibinfo{howpublished}{\url{https://www.3dcadworld.com/the-failed-promise-of-parametric-cad/}}.
\newblock
\newblock
\shownote{(Accessed on 09/06/2019).}


\end{thebibliography}

\end{document}